\documentclass[acmsmall,screen,nonacm]{acmart}

\usepackage{amsmath,amssymb}
\usepackage{placeins}
\usepackage{float}
\usepackage{longtable,booktabs,array}
\usepackage{calc}
\usepackage{etoolbox}
\AtBeginEnvironment{longtable}{\sloppy}
\makeatletter
\patchcmd\longtable{\par}{\if@noskipsec\mbox{}\fi\par}{}{}
\makeatother
\makeatletter
\newsavebox\pandoc@box
\newcommand*\pandocbounded[1]{%
  \sbox\pandoc@box{#1}%
  \Gscale@div\@tempa{\textheight}{\dimexpr\ht\pandoc@box+\dp\pandoc@box\relax}%
  \Gscale@div\@tempb{\linewidth}{\wd\pandoc@box}%
  \ifdim\@tempb\p@<\@tempa\p@\let\@tempa\@tempb\fi%
  \ifdim\@tempa\p@<\p@\scalebox{\@tempa}{\usebox\pandoc@box}%
  \else\usebox\pandoc@box%
  \fi%
}
\makeatother
\providecommand{\tightlist}{\setlength{\itemsep}{0pt}\setlength{\parskip}{0pt}}
\AtBeginDocument{}

\acmJournal{CSUR}
\setcopyright{none}
\numberwithin{table}{section}

\begin{document}

\title{Towards Safe World Models: Risks, Defenses, and Evaluation}

% Names are romanized for the English-language ACM metadata. Add official
% emails, ORCID identifiers, and corresponding-author marks before submission.
\author{Xiaoma Wang}
\email{xiaomawang@bupt.edu.cn}
\affiliation{\institution{Beijing University of Posts and Telecommunications}\city{Beijing}\country{China}}
\author{Litian Zhang}
\email{litianzhang@bupt.edu.cn}
\correspondingauthor
\affiliation{\institution{Beijing University of Posts and Telecommunications}\city{Beijing}\country{China}}
\author{Xinze Shi}
\affiliation{\institution{Beijing University of Posts and Telecommunications}\city{Beijing}\country{China}}
\author{Zhengxuan Hu}
\affiliation{\institution{Beijing University of Posts and Telecommunications}\city{Beijing}\country{China}}
\author{Zihan Yang}
\affiliation{\institution{Beijing University of Posts and Telecommunications}\city{Beijing}\country{China}}
\author{Mingqi Ren}
\affiliation{\institution{Beijing University of Posts and Telecommunications}\city{Beijing}\country{China}}
\author{You Wu}
\affiliation{\institution{Beijing University of Posts and Telecommunications}\city{Beijing}\country{China}}
\author{Bingyu Yan}
\affiliation{\institution{Beihang University}\city{Beijing}\country{China}}
\author{Zelong Yu}
\affiliation{\institution{Beihang University}\city{Beijing}\country{China}}
\author{Zhenjie Su}
\affiliation{\institution{Beijing University of Posts and Telecommunications}\city{Beijing}\country{China}}
\author{Yuming Shang}
\email{shangym@bupt.edu.cn}
\affiliation{\institution{Beijing University of Posts and Telecommunications}\city{Beijing}\country{China}}
\author{Xi Zhang}
\email{zhangx@bupt.edu.cn}
\affiliation{\institution{Beijing University of Posts and Telecommunications}\city{Beijing}\country{China}}
\author{Chaozhuo Li}
\email{lichaozhuo@bupt.edu.cn}
\affiliation{\institution{Beijing University of Posts and Telecommunications}\city{Beijing}\country{China}}
\author{Philip S. Yu}
\email{psyu@uic.edu}
\affiliation{\institution{University of Illinois Chicago}\city{Chicago}\state{Illinois}\country{USA}}
\renewcommand{\shortauthors}{Wang et al.}

\begin{abstract}
World models support planning, control, learning from imagined experience, and model-mediated evaluation by predicting action-conditioned consequences. Their safety cannot be inferred from predictive accuracy alone: errors in observations, state and memory, dynamics, evaluation, execution, or feedback can propagate into unsafe decisions and persist through updates. This work organizes the literature on world-model safety around a six-stage lifecycle covering data and observations, state and memory, dynamics and imagination, evaluation and planning, action and control, and feedback and update. We synthesize non-adversarial failures, attacks, defenses, evaluation protocols, benchmarks, and deployment conditions, and distinguish evidence directness from outcome level. The current evidence corpus contains 163 cited works. We use the framework to identify where claims are supported, where attribution remains unresolved, and which measurements are needed to connect model-level behavior to closed-loop and operational safety.
\end{abstract}

% Provisional entries. Regenerate the CCS block with the ACM CCS tool before submission.

\maketitle
\clearpage

\FloatBarrier
\section{Introduction}\label{introduction}

World models enable agents to anticipate the consequences of their actions before acting. By learning how an environment changes, they support planning, policy learning from imagined experience, and simulation. Their use spans latent dynamics for control, physical robot learning, and interactive generative environments \citep{hafner2020dreamer, wu2023daydreamer, bruce2024genie}. As these predictions influence decisions in the physical world, their errors acquire safety consequences: an inaccurate future can make an unsafe action appear acceptable.

The difficulty is that prediction quality and safe behavior are connected through a decision process. Model errors can accumulate over imagined trajectories, and optimization can favor actions whose outcomes the model predicts too optimistically \citep{janner2019mbpo, yu2020mopo}. Consider a robot approaching an occluded obstacle. If its internal state loses the obstacle's location, the model may predict a clear path, the planner may select it, and execution delay may leave insufficient time to stop. This hypothetical example illustrates how a local error can cross several functional boundaries before causing harm. The resulting observations may also affect subsequent state estimates or learning. Safety analysis must therefore follow predictions through the decisions and feedback they influence.

These deviations can arise without an attacker, through incomplete data, partial observability, model bias, or unsuitable objectives. They can also result from deliberate manipulation of data, observations, or other system interfaces \citep{brunke2022safelearning, biggio2018wildpatterns}. Distinguishing these causes is necessary for attribution, while tracing their propagation requires a shared view of the system.

We focus on learned models that predict action-conditioned environmental consequences used in learning, planning, simulation, or safety assessment. A \emph{world-model-centric system} includes the components that construct, consume, execute, and update these predictions. Our scope covers non-adversarial failures and attacks, their defenses, and the evidence used to assess safety. Adjacent perception or control research is included when it explains an interface through which world-model risk can arise or propagate; generic prediction accuracy alone does not establish relevance to a safety claim. Section 2 develops the functional definition and architectural background.

Prior work offers complementary foundations. Model-based reinforcement learning is organized around model learning and use \citep{moerland2023modelbased}, while safe learning in robotics emphasizes constraints, control, and guarantees \citep{brunke2022safelearning}. Within ACM Computing Surveys, \citet{mohseni2023mlsafety} connect ML dependability techniques to engineering safety strategies, and \citet{ashmore2021assuring} organize assurance around the ML lifecycle. VLA safety literature brings embodied threats, defenses, and evaluation into a common discussion \citep{li2026vlasafety}. Our organizing question is more specific: how does a deviation in a learned account of the environment become an unsafe decision, and what evidence supports that connection? This focus links the internal predictive process to its consequences across different world-model architectures.

Deployment-oriented agent research supplies a complementary interface
perspective. The architecture-lifecycle account of computer-use agents
connects perception and decisions to tool permissions, execution, and
maintenance \citep{chen2026cuareliability}. These software-agent
dependencies help identify where a world-model prediction may acquire
action authority, while the present analysis additionally requires
action-conditioned dynamics and a trace from prediction to consequence.

To answer this question, we follow the information needed to make and revise a model-informed decision. Experience and observations must first be acquired and interpreted as a current state. That state supports predictions of possible futures; objectives and constraints then determine which futures guide action. Execution changes the environment, and the resulting feedback revises subsequent decisions and, where applicable, the model itself. This reasoning yields six functional stages: \textbf{Data and Observations (S1)}, \textbf{State and Memory (S2)}, \textbf{Dynamics and Imagination (S3)}, \textbf{Evaluation and Planning (S4)}, \textbf{Action and Control (S5)}, and \textbf{Feedback and Update (S6)}. Separating observations from state exposes information loss and stale memory; separating prediction from evaluation distinguishes incorrect futures from unsafe selection; separating execution from feedback identifies how deviations persist through interaction and updates. The stages describe dependencies rather than six mandatory software modules, and they need not all operate at the same frequency.

\begin{figure}[!htbp]
\centering
\includegraphics[width=\linewidth,height=0.36\textheight,keepaspectratio]{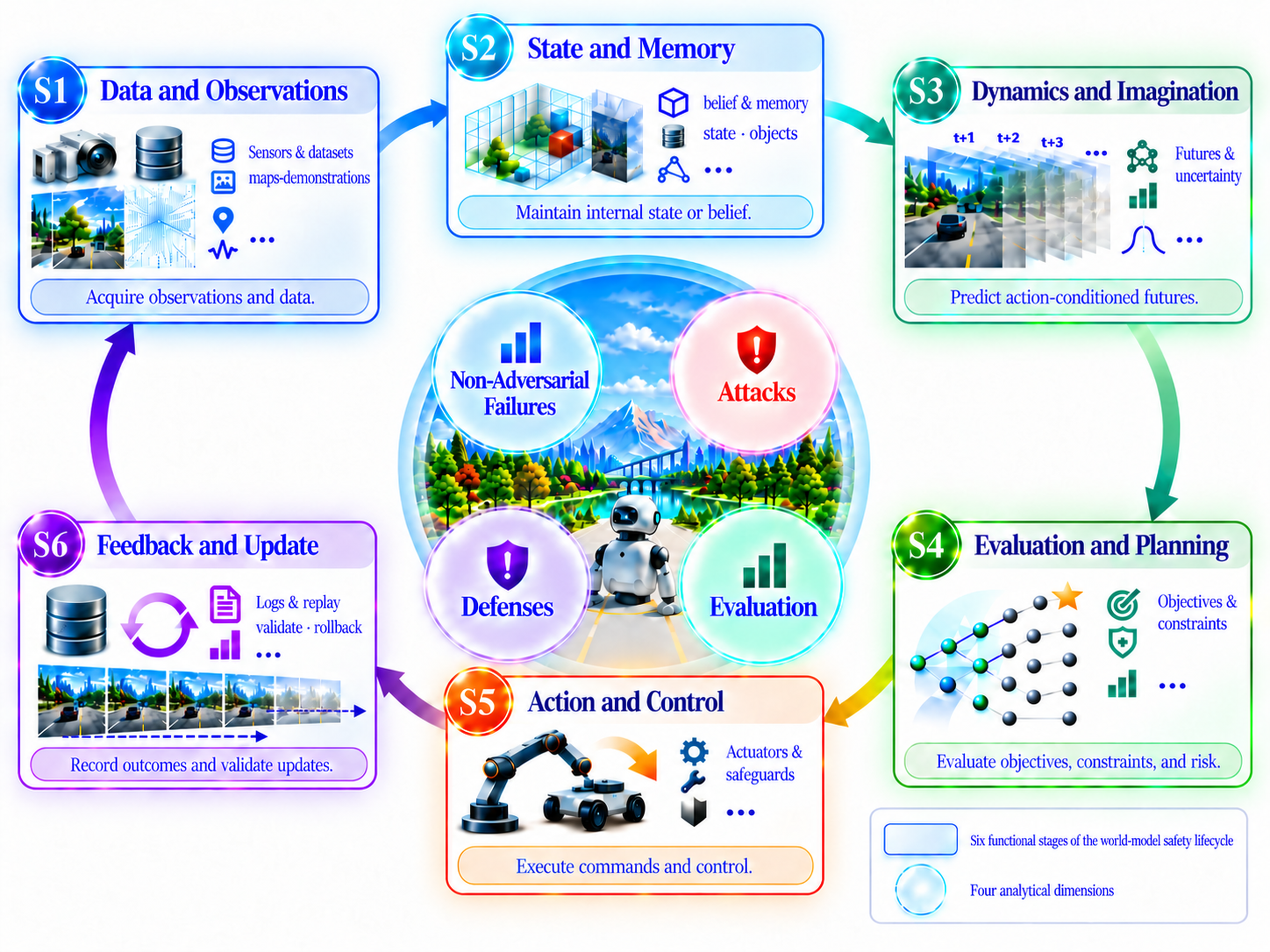}
\caption{Six-stage world-model safety loop and four analytical dimensions.}
\Description{A six-stage loop links data and observations, state and memory, dynamics and imagination, evaluation and planning, action and control, and feedback and update. Surrounding labels identify failures, attacks, defenses, and evaluation.}
  \label{fig:safety-loop}
\end{figure}

Table~\ref{tab:lifecycle-orientation} provides a compact orientation map. Its rows are four analytical dimensions---non-adversarial failures, attacks, defenses, and evaluation---that cut across all six stages. Each cell names one representative work; the evaluation dimension groups model-level resources across S1--S3 because those resources often span state, prediction, and uncertainty rather than a single stage.

\small\begin{longtable}[]{@{}>{\raggedright\arraybackslash}p{(\linewidth - 45pt) * \real{0.18}}@{\hspace{6pt}}>{\raggedright\arraybackslash}p{(\linewidth - 45pt) * \real{0.1366}}@{\hspace{6pt}}>{\raggedright\arraybackslash}p{(\linewidth - 45pt) * \real{0.1366}}@{\hspace{6pt}}>{\raggedright\arraybackslash}p{(\linewidth - 45pt) * \real{0.1366}}@{\hspace{6pt}}>{\raggedright\arraybackslash}p{(\linewidth - 45pt) * \real{0.1366}}@{\hspace{6pt}}>{\raggedright\arraybackslash}p{(\linewidth - 45pt) * \real{0.1366}}@{\hspace{6pt}}>{\raggedright\arraybackslash}p{(\linewidth - 45pt) * \real{0.1366}}@{}}
\caption{Representative works across the six-stage safety lifecycle.}\label{tab:lifecycle-orientation}\\
\toprule\noalign{}
\begin{minipage}[b]{\linewidth}\raggedright
Analytical dimension
\end{minipage} & \begin{minipage}[b]{\linewidth}\raggedright
\parbox[t]{\linewidth}{\raggedright S1\\Data\\and\\Observations}
\end{minipage} & \begin{minipage}[b]{\linewidth}\raggedright
\parbox[t]{\linewidth}{\raggedright S2\\State\\and\\Memory}
\end{minipage} & \begin{minipage}[b]{\linewidth}\raggedright
\parbox[t]{\linewidth}{\raggedright S3\\Dynamics\\and\\Imagination}
\end{minipage} & \begin{minipage}[b]{\linewidth}\raggedright
\parbox[t]{\linewidth}{\raggedright S4\\Evaluation\\and\\Planning}
\end{minipage} & \begin{minipage}[b]{\linewidth}\raggedright
\parbox[t]{\linewidth}{\raggedright S5\\Action\\and\\Control}
\end{minipage} & \begin{minipage}[b]{\linewidth}\raggedright
\parbox[t]{\linewidth}{\raggedright S6\\Feedback\\and\\Update}
\end{minipage} \\
\midrule\noalign{}
\endfirsthead
\toprule\noalign{}
\begin{minipage}[b]{\linewidth}\raggedright
Analytical dimension
\end{minipage} & \begin{minipage}[b]{\linewidth}\raggedright
\parbox[t]{\linewidth}{\raggedright S1\\Data\\and\\Observations}
\end{minipage} & \begin{minipage}[b]{\linewidth}\raggedright
\parbox[t]{\linewidth}{\raggedright S2\\State\\and\\Memory}
\end{minipage} & \begin{minipage}[b]{\linewidth}\raggedright
\parbox[t]{\linewidth}{\raggedright S3\\Dynamics\\and\\Imagination}
\end{minipage} & \begin{minipage}[b]{\linewidth}\raggedright
\parbox[t]{\linewidth}{\raggedright S4\\Evaluation\\and\\Planning}
\end{minipage} & \begin{minipage}[b]{\linewidth}\raggedright
\parbox[t]{\linewidth}{\raggedright S5\\Action\\and\\Control}
\end{minipage} & \begin{minipage}[b]{\linewidth}\raggedright
\parbox[t]{\linewidth}{\raggedright S6\\Feedback\\and\\Update}
\end{minipage} \\
\midrule\noalign{}
\endhead
\bottomrule\noalign{}
\endlastfoot
Non-adversarial failure & WAKER \citep{rigter2024waker} & PIGDreamer \citep{huang2025pigdreamer} & Calibrated MBRL \citep{malik2019calibrated} & CAP \citep{ma2022cap} & Recovery RL \citep{thananjeyan2021recovery} & Continual updates \citep{kessler2023continual, liu2025continual} \\
Attack & BadDreamer \citep{Shuai2026BadDreamer} & ARB4WM \citep{Zhang2026ARB4WM} & DINO-WM/PPO \citep{Rathbun2026TargetingWM} & AdvDO \citep{Cao2022AdvDO} & BadWAM \citep{Li2026BadWAM} & TrojanWorld \citep{Huang2026TrojanWorld} \\
Defense & ReSim \citep{yang2025resim} & RePo \citep{zhu2023repo} & MOPO \citep{yu2020mopo} & SafeDreamer \citep{huang2024safedreamer} & Shielding \citep{alshiekh2018shielding} & AdaWM \citep{wang2025adawm} \\
Evaluation & \multicolumn{3}{c}{Calibrated MBRL \citep{malik2019calibrated}} & SafeDreamer \citep{huang2024safedreamer} & SafeBench \citep{xu2022safebench} & SAFE-SMART / ASIMOV \citep{sakano2025safesmart, sermanet2025asimov} \\
\end{longtable}

This work provides five connected contributions: (1) a functional scope and architectural account of world models as components of decision-making systems; (2) a six-stage framework for locating safety-relevant deviations and tracing their propagation; (3) an organization of non-adversarial failures and attacks that preserves their different causal assumptions; (4) a synthesis of defenses and evaluation evidence, distinguishing model, planning, behavior, and operational outcomes; and (5) a comparison of deployment conditions and the research questions they raise for world-model safety. Section 2 introduces world models and their evolution. Section 3 presents the evidence methodology and safety framework. Sections 4--5 examine failures and attacks, Section 6 examines defenses, and Section 7 addresses evaluation and benchmarks. Sections 8--9 discuss deployment domains and open research problems. Section 10 concludes the work.

\FloatBarrier
\section{Foundations and Evolution of World Models}\label{foundations-and-evolution-of-world-models}

This section establishes what counts as a world model, how major architectural families represent and use environmental dynamics, and how the field has evolved. It keeps safety analysis at the level needed to motivate later distinctions; the systematic treatment of safety objects, failure propagation, threats, defenses, and evidence begins with the six-stage framework in Section 3.

\subsection{Problem Setting and Functional Definition}\label{problem-setting-and-functional-definition}

Consider an environment with latent state \(s_t\), observation \(o_t\), action \(a_t\), reward \(r_t\), safety cost \(c_t\), and transition process \(p(s_{t+1}\mid s_t,a_t)\). In a partially observable setting, an agent acts from a history \(h_t=(o_{\leq t},a_{<t})\) rather than from the true state. A world model constructs an internal state or belief

\[
z_t = e_{\theta}(h_t)
\]

and an action-conditioned predictive distribution

\[
p_{\theta}(z_{t+1},o_{t+1},r_t,c_t,d_t\mid z_t,a_t),
\]

where \(d_t\) may represent termination, event, or constraint-related variables. The predictive distribution may be factorized into aleatoric uncertainty (environmental stochasticity) and epistemic uncertainty (model uncertainty). A planner can roll this distribution forward under a candidate action sequence \(a_{t:t+H-1}\), while an imagination-based learner can use sampled or expected rollouts to update a policy or value function. The same functional model may be deterministic or probabilistic, reconstruct observations or omit them, and predict one step or an extended horizon.

This formulation reflects the partially observable nature of most relevant systems. The internal state is not assumed to equal the physical state; it is a learned statistic of history whose adequacy depends on the downstream query. Two histories may be equivalent for reward prediction yet differ in a variable needed to choose between safe actions. This distinction between task sufficiency and safety sufficiency motivates the protected-object analysis in Section 3, but does not require a separate definition of the model.

Throughout this work, \textbf{world-model-centric} refers to systems within the scope set out in the Introduction and formalized here. The category includes learned dynamics used in model-predictive control, recurrent state-space models used for latent imagination, object- or occupancy-centric predictors used for temporal reasoning, and interactive generative environments used to train or evaluate agents. Static encoders and one-shot predictors remain outside the main scope unless their outputs become the state of a temporal decision process.

\begin{figure}[!htbp]
\centering
\includegraphics[width=\linewidth,height=0.36\textheight,keepaspectratio]{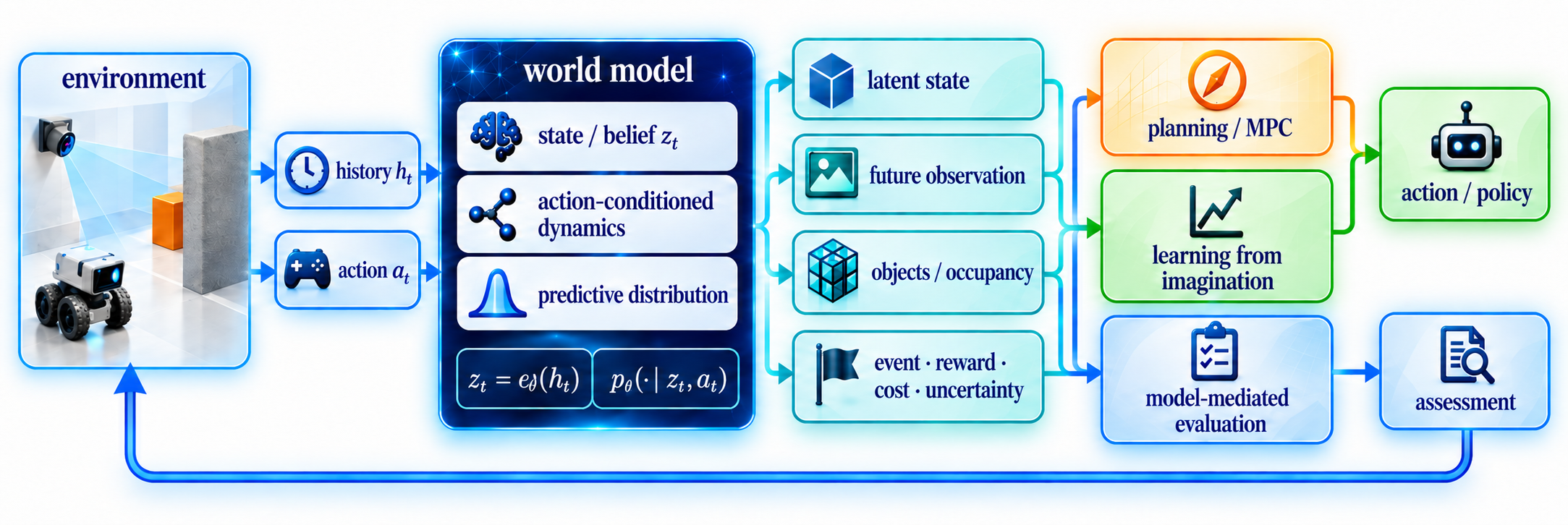}
\caption{Functional world-model interface.}
\Description{Observation and action history enter an internal state representation and an action-conditioned predictive model. The outputs support planning, imagined learning, or safety assessment.}
  \label{fig:functional-interface}
\end{figure}

\subsection{Architectural Families and Interfaces}\label{architectural-families-and-interfaces}

Architectural labels change quickly, but safety-relevant interfaces are comparatively stable. We group world models into five non-exclusive families.

\paragraph{Latent state-space models}\label{latent-state-space-models}

Latent state-space models encode observations into a compact state and learn recurrent action-conditioned dynamics. PlaNet introduced planning in a learned latent dynamics model from pixels, while Dreamer learned behavior from latent imagined trajectories \citep{hafner2019planet, hafner2020dreamer}. Later systems improved discrete representation, scale, and generality \citep{hafner2021dreamerv2, hafner2025dreamerv3}. Their main advantage is efficient prediction and learning in a compact space; their main interpretive limitation is that the physical and semantic content retained by the latent state may not be directly visible.

\paragraph{Value-equivalent and task-oriented models}\label{value-equivalent-and-task-oriented-models}

Some models predict only what is needed for planning toward a task objective. MuZero learns an internal transition, policy, reward, and value representation suitable for tree search without reconstructing observations \citep{schrittwieser2020muzero}. TD-MPC combines learned latent dynamics with terminal value estimation and model-predictive control \citep{hansen2022tdmpc}, and TD-MPC2 extends this design to multi-task continuous control \citep{hansen2024tdmpc2}. These models trade reconstructive completeness for decision utility, so what they preserve is determined largely by rewards, tasks, and search targets.

\paragraph{Observation-generative and interactive models}\label{observation-generative-and-interactive-models}

Observation-generative models predict future images, video, or other high-dimensional observations. Interactive generative environments additionally condition on user or agent actions and support counterfactual trajectories. Genie demonstrates action-controllable generative environments learned from unlabeled video \citep{bruce2024genie}, while RoboDreamer generates compositional visual futures for robot tasks \citep{zhou2024robodreamer}. Their explicit outputs support inspection and broad scene modeling, although perceptual plausibility need not imply correct causal or physical dynamics.

\paragraph{Object-centric and spatial world models}\label{object-centric-and-spatial-world-models}

Object-, occupancy-, bird's-eye-view-, and three-dimensional scene models represent entities and spatial relations relevant to interaction. OccWorld, for example, predicts future three-dimensional occupancy for autonomous driving \citep{zheng2024occworld}. These representations make geometry and reachable regions more explicit than opaque latent states, but depend on detection, tracking, data association, coordinate transformation, occlusion handling, and discretization.

\paragraph{Hybrid world-action and evaluator models}\label{hybrid-world-action-and-evaluator-models}

In increasingly integrated systems, the model jointly predicts future states and actions, shares parameters with a policy, or evaluates candidate behavior. Integration may improve learning and long-horizon coherence, while making the boundary among representation, prediction, evaluation, and action less explicit. Such systems are classified by their functional dependencies rather than by the visible neural-module boundary.

Table~\ref{tab:architectural-families} contrasts the five families by the predictions they expose, the decisions they support, and the safety-relevant information they may omit.

\begin{longtable}[]{@{}
  >{\raggedright\arraybackslash}p{(\linewidth - 6\tabcolsep) * \real{0.2500}}
  >{\raggedright\arraybackslash}p{(\linewidth - 6\tabcolsep) * \real{0.2500}}
  >{\raggedright\arraybackslash}p{(\linewidth - 6\tabcolsep) * \real{0.2500}}
  >{\raggedright\arraybackslash}p{(\linewidth - 6\tabcolsep) * \real{0.2500}}@{}}
\caption{Architectural families and their safety-relevant implications.}\label{tab:architectural-families}\\
\toprule\noalign{}
\begin{minipage}[b]{\linewidth}\raggedright
Family
\end{minipage} & \begin{minipage}[b]{\linewidth}\raggedright
Typical prediction
\end{minipage} & \begin{minipage}[b]{\linewidth}\raggedright
Decision role
\end{minipage} & \begin{minipage}[b]{\linewidth}\raggedright
Safety-relevant implication
\end{minipage} \\
\midrule\noalign{}
\endfirsthead
\toprule\noalign{}
\begin{minipage}[b]{\linewidth}\raggedright
Family
\end{minipage} & \begin{minipage}[b]{\linewidth}\raggedright
Typical prediction
\end{minipage} & \begin{minipage}[b]{\linewidth}\raggedright
Decision role
\end{minipage} & \begin{minipage}[b]{\linewidth}\raggedright
Safety-relevant implication
\end{minipage} \\
\midrule\noalign{}
\endhead
\bottomrule\noalign{}
\endlastfoot
Latent state-space & Latent transition, reward, continuation & Planning or imagined policy learning & Hazard information may be compressed, aliased, or difficult to audit \\
Value-equivalent/task-oriented & Reward, value, policy-relevant latent state & Search and model-predictive control & Task sufficiency may omit safety-relevant facts or alternative outcomes \\
Observation-generative/interactive & Images, video, events, action-conditioned futures & Simulation, planning, data generation & Perceptual realism may conceal causal and physical error \\
Object-centric/spatial & Objects, trajectories, occupancy, geometry & Motion reasoning and constraint checking & Rare objects, occlusion, and multimodal futures may be missed \\
Hybrid world-action/evaluator & Future and action, or future and score & Integrated control or policy evaluation & Shared representations and data create common-mode failure \\
\end{longtable}

These families overlap: one system may contain a latent transition, a visual decoder, an occupancy head, and a policy head. Family labels describe architecture and representation only. Section 3 separately defines the loop stage, protected object, outcome level, and evidence directness used for safety analysis.

\subsection{Modes of Model Use in Decision Making}\label{modes-of-model-use-in-decision-making}

Architecture does not determine how a world model affects decisions. We distinguish three principal modes of use.

In \textbf{planning with a model}, an optimizer, tree search, or model-predictive controller queries the learned transition to compare candidate action sequences. The model remains inside the deployment-time decision loop, and the planner's search distribution determines which parts of the learned dynamics matter most.

In \textbf{learning from imagination}, generated trajectories provide targets or gradients for a policy or value function. The resulting policy may no longer query the model at deployment time, but it can retain regularities and biases learned from imagined experience.

In \textbf{model-mediated evaluation}, the model simulates candidate policies, generates test situations, predicts events, or ranks behavior without directly issuing actions. Its evidential value depends on whether the evaluator is sufficiently independent of the system being evaluated and whether its simulated conditions cover the relevant operating domain.

These three modes describe how predictions are consumed, not where a safety deviation originates. Section 3 supplies the six-stage reference architecture used to trace state construction, prediction, evaluation, action, and feedback across all three modes.

\subsection{Evolution of World Models}\label{evolution-of-world-models}

World models have evolved through changes in representation, prediction horizon, model use, and deployment setting. Early integrated learning-and-planning systems established that experience could update an internal model and that simulated experience could update behavior \citep{sutton1990dyna}. Probabilistic dynamics models later improved sample-efficient continuous control by representing model uncertainty and planning through ensembles \citep{chua2018pets}.

Latent dynamics made high-dimensional observation streams tractable for online planning. PlaNet demonstrated that a compact recurrent state-space model could support planning directly from pixels \citep{hafner2019planet}. Dreamer and its successors then used large numbers of latent trajectories to train policy and value functions \citep{hafner2020dreamer, hafner2021dreamerv2, hafner2025dreamerv3}, expanding the model's role from online prediction to imagined experience generation.

Task-oriented planning models and scalable control systems broadened the modeled domains and objectives. MuZero showed that a useful planning model need not reproduce observations \citep{schrittwieser2020muzero}, while TD-MPC2 demonstrated a scalable multi-task world model for continuous control \citep{hansen2024tdmpc2}. These systems strengthened the distinction between reconstructing the environment and preserving information useful for decisions.

World models have also moved toward physical and open-ended interaction. DayDreamer applied world-model learning on physical robots \citep{wu2023daydreamer}, spatial models predict future occupancy in driving environments \citep{zheng2024occworld}, and generative interactive models synthesize action-conditioned visual worlds \citep{bruce2024genie, zhou2024robodreamer}. The modeled state has consequently expanded from compact control variables to multimodal scenes, objects, events, and interactive visual futures.

Across this evolution, predictions are used by more components, over longer horizons, and in settings with progressively richer observations and interactions. These changes motivate the lifecycle analysis in Section 3. The historical sequence itself does not establish improved safety; it identifies the model capabilities and dependencies that the six-stage framework must analyze.

\subsection{Boundaries with Neighboring Concepts}\label{boundaries-with-neighboring-concepts}

The functional definition also sets boundaries with simulators, digital twins, policies, value functions, and foundation models. These terms overlap with world models but are not interchangeable.

A \textbf{simulator} is an executable transition mechanism and may be constructed from physical equations, rules, human design, recorded data, or learned components. A learned simulator used to predict decision consequences may satisfy our functional definition of a world model. A world model, however, may predict only a compact latent state and need not provide a complete environment interface. The distinction matters for assurance because a physics simulator may carry explicit validity assumptions, whereas a learned simulator carries data and generalization assumptions.

A \textbf{digital twin} is typically tied to a particular physical asset or process and emphasizes synchronization between the digital representation and its real counterpart. A world model may instead learn transferable regularities across assets and tasks. When the two are combined, synchronization, identity, version consistency, and domain validity become additional safety objects.

A \textbf{policy} maps information states to actions. A world model maps current state and action to possible consequences. End-to-end systems may share parameters, but functional separation remains necessary for diagnosis: unsafe behavior can originate in state inference, predicted dynamics, objective specification, planning, or action decoding. Treating the entire network as a policy conceals these distinct intervention points.

A \textbf{value function} compresses expected future return into a scalar or distribution. A world model exposes an iterable structure through which alternative action-conditioned futures can be evaluated. Value-equivalent models blur the structural boundary, but not the functional one. If an internal transition is repeatedly used by search to reason about action consequences, it is within scope; a standalone value estimate without such a transition is not independently treated as a world model.

A \textbf{foundation model} is characterized primarily by scale, broad pretraining, and adaptability. It may contain or support a world model, but broad generative competence does not by itself establish action-conditioned dynamics or a decision-loop role. This work includes a foundation model only to the extent that its internal or external predictive process satisfies the functional criterion above.

These boundaries support attribution rather than exclusion. Section 3 uses them to identify which component is evaluated, which decision path consumes its predictions, and how far the resulting evidence can support a safety claim.

\FloatBarrier
\section{Analytical Framework for World-Model Safety}\label{analytical-framework-for-world-model-safety}

This section defines how world-model safety is analyzed throughout this work. It first specifies the evidence corpus and the result-level coding protocol, then decomposes a world-model-centric system into a six-stage lifecycle. The subsequent analysis follows the same sequence: identify the protected object, trace how a deviation propagates toward harm, characterize the threat or operating condition, and determine what level of evidence supports the claim. The framework is analytical rather than architectural: its stages identify causal dependencies and intervention points, not mandatory software modules.

\subsection{Evidence Corpus and Coding Protocol}\label{evidence-corpus-and-coding-protocol}

We constructed the evidence corpus through a structured, claim-centered process rather than a frequency-based meta-analysis. The search scope covered work available through September 2026 on learned world models, model-based or latent dynamics, action-conditioned predictive environments, interactive generative environments, and world-model-enabled planning, control, monitoring, or safety assessment. Boolean query blocks combined these concepts with safety terms: (``world model'' OR ``learned dynamics'' OR ``latent dynamics'' OR ``predictive environment'') AND (safety OR robustness OR uncertainty OR attack OR defense OR assurance OR evaluation OR benchmark OR monitoring OR recovery OR deployment), with additional embodied, robotics, navigation, and driving terms where the application context was relevant. We searched bibliographic indexes and publisher repositories for published work and consulted preprint and open-publication repositories when a rapidly emerging topic had not yet acquired a stable archival record. The bibliography records the publication status of each source; recent preprints are retained only when they provide direct evidence or a necessary interface for the safety argument.

Studies were included when they satisfied at least one of the following conditions: (i) they evaluated an action-conditioned environmental predictor or its learned state, rollout, or uncertainty; (ii) they measured how such predictions affected planning, policy learning, execution, monitoring, or updating; (iii) they supplied a safety-relevant neighboring interface whose assumptions could transmit world-model risk; or (iv) they provided an evaluation resource, benchmark, formal method, or synthesis needed to interpret a world-model safety claim. We excluded static perception or one-shot generation without a temporal decision role, generic task-performance papers with no relevant predictive or safety interface, duplicate versions of the same result, and commentary without a technical object or evaluable protocol. A source could therefore be included without being direct world-model evidence, but its role had to be coded explicitly.

Screening and synthesis were performed at the level of claims and reported results rather than assigning one permanent label to an entire paper. For each included result, we recorded the primary loop stage, risk mechanism, protected object, defense function where applicable, evaluation outcome level, domain and operating assumptions, evidence directness (E1--E4), and publication status. The same result retains the same coding wherever it is cited; distinct results from one paper may receive different codes. This procedure is formalized in Section 3.7 and is intended to prevent capability evidence, interface evidence, and end-to-end safety evidence from being conflated.

A supplementary interface-focused search examined language-agent input,
tool, memory, runtime, and evaluation mechanisms. These additional
sources are coded as E4 unless a specific learned-world-model dependency
is tested; their inclusion does not broaden the main scope to generic
language-agent safety.

The resulting corpus contains \textbf{163 distinct cited works} in the present version of this work. These records span foundational world-model research, safety and assurance methods, attacks and defenses, evaluation resources, and deployment-oriented studies. Because a work may support more than one section or evidence role, section-level citation counts are not additive. The merged source materials did not preserve per-database hit counts or a PRISMA-style screening flow, so we report the deduplicated evidence corpus and its coding rules rather than inventing retrieval statistics. The corpus should be read as a transparent, structured evidence base rather than as an exhaustive census of every paper using the term \emph{world model}. The main limitations are vocabulary drift, uneven publication status in this fast-moving area, and the scarcity of studies that expose the complete prediction-to-consequence chain.

\subsection{Lifecycle Decomposition: The Six-Stage Safety Loop}\label{lifecycle-decomposition-the-six-stage-safety-loop}

The six-stage lifecycle decomposition is the organizing backbone of this work. Section 2 describes what a world model is and how it participates in decision making; the stages below identify where failures and attacks enter, where defenses intervene, and where evaluation evidence is collected. Sections 4--8 use the same stage names and ordering.

We abstract a world-model-centric system into six functional stages that form a forward decision path and a feedback path.

We denote the stages as \textbf{S1 Data and Observations}, \textbf{S2 State and Memory}, \textbf{S3 Dynamics and Imagination}, \textbf{S4 Evaluation and Planning}, \textbf{S5 Action and Control}, and \textbf{S6 Feedback and Update}. The decomposition is functional rather than architectural. One neural network may implement several stages, and one stage may be distributed across multiple components. The purpose is to expose interfaces through which error, attack, evidence, and defensive responsibility can be traced.

The six stages are a minimum causal decomposition, not a claim that every implementation has six modules. Objective and safety specification are represented within Evaluation and Planning because they determine admissibility and ranking, while human oversight, communication, and the physical environment are treated as boundary conditions on Action and Control and Feedback and Update. If a study models any of these elements explicitly, it receives an additional interface annotation rather than forcing them into a separate architectural family.

\textbf{S1 Data and Observations} includes offline datasets, demonstrations, maps, labels, synthetic experience, and online sensor streams. The stage concerns both information content and provenance. Coverage, integrity, synchronization, and modality agreement determine what the system can infer and learn.

\textbf{S2 State and Memory} transforms current and historical information into an internal state. It includes latent state estimation, belief tracking, object and scene representation, retrieval, episodic memory, and temporal context. Its central safety responsibility is to preserve decision-relevant distinctions despite partial observability and noise.

\textbf{S3 Dynamics and Imagination} predicts action-conditioned futures. The term \emph{dynamics} denotes transition structure, while \emph{imagination} denotes the use of that structure to generate hypothetical rollouts. Depending on the model, outputs may be latent states, observations, objects, occupancy, events, rewards, costs, or uncertainty distributions.

\textbf{S4 Evaluation and Planning} assigns value, cost, risk, or feasibility to predicted futures and selects a plan or policy update. It includes objective specification, constraint modeling, candidate generation, ranking, search, and risk aggregation.

\textbf{S5 Action and Control} converts the selected plan into commands and environmental intervention. It includes action decoding, timing, low-level tracking, saturation, safety filtering, backup control, and emergency behavior. Shielding is one concrete instance of this stage: a monitor can intercept or correct an unsafe action against an explicit specification \citep{alshiekh2018shielding}. Although some mechanisms are external to the learned model, they determine whether an upstream model error becomes a physical consequence.

\textbf{S6 Feedback and Update} records outcomes and changes later inference or learning. It includes logs, replay buffers, memory refresh, online adaptation, model and policy updates, validation gates, deployment versions, and rollback. This stage closes the loop: consequences become future evidence, but corrupted or unrepresentative evidence can also make a failure persistent.

The forward direction explains hazard formation. The feedback direction explains correction, persistence, and amplification. A safety analysis that stops at action selection misses contaminated logging, continual-learning attacks, and post-update regression. An analysis that focuses only on data governance misses situations in which a well-curated model is exploited by planning or fails beyond its trusted horizon.

\subsection{Protected Objects and Safety Properties}\label{protected-objects-and-safety-properties}

Table~\ref{tab:six-stage-map} summarizes the compact analytical map used throughout this work. It keeps the protected object, dominant failure patterns, and minimum evidence visible without repeating the full diagnostic questions from the surrounding text.

\begin{longtable}[]{@{}
  >{\raggedright\arraybackslash}p{(\linewidth - 8\tabcolsep) * \real{0.2000}}
  >{\raggedright\arraybackslash}p{(\linewidth - 8\tabcolsep) * \real{0.2000}}
  >{\raggedright\arraybackslash}p{(\linewidth - 8\tabcolsep) * \real{0.2000}}
  >{\raggedright\arraybackslash}p{(\linewidth - 8\tabcolsep) * \real{0.2000}}
  >{\raggedright\arraybackslash}p{(\linewidth - 8\tabcolsep) * \real{0.2000}}@{}}
\caption{Compact six-stage map for world-model safety.}\label{tab:six-stage-map}\\
\toprule\noalign{}
\begin{minipage}[b]{\linewidth}\raggedright
Stage
\end{minipage} & \begin{minipage}[b]{\linewidth}\raggedright
Protected object
\end{minipage} & \begin{minipage}[b]{\linewidth}\raggedright
Failure type
\end{minipage} & \begin{minipage}[b]{\linewidth}\raggedright
Representative examples
\end{minipage} & \begin{minipage}[b]{\linewidth}\raggedright
Minimum evidence
\end{minipage} \\
\midrule\noalign{}
\endfirsthead
\toprule\noalign{}
\begin{minipage}[b]{\linewidth}\raggedright
Stage
\end{minipage} & \begin{minipage}[b]{\linewidth}\raggedright
Protected object
\end{minipage} & \begin{minipage}[b]{\linewidth}\raggedright
Failure type
\end{minipage} & \begin{minipage}[b]{\linewidth}\raggedright
Representative examples
\end{minipage} & \begin{minipage}[b]{\linewidth}\raggedright
Minimum evidence
\end{minipage} \\
\midrule\noalign{}
\endhead
\bottomrule\noalign{}
\endlastfoot
\textbf{S1 -- Data and Observations} & Coverage, provenance, timing, modality consistency & Omission, drift, delay, poisoning & Long-tail scenes; sensor mismatch & Coverage, provenance, and sensitivity checks \\
\textbf{S2 -- State and Memory} & Hazard state, identity, relations, temporal context & Aliasing, compression, stale belief, false grounding & Occlusion; memory loss; state trigger & State probes, hazard recall, intervention \\
\textbf{S3 -- Dynamics and Imagination} & Futures, reachable events, costs, uncertainty & Bias, rollout error, missed branches, miscalibration & Long-horizon rollout; rare hazard & Stratified prediction and decision sensitivity \\
\textbf{S4 -- Evaluation and Planning} & Costs, constraints, feasibility, ranking & Omission, reversal, exploitation, score manipulation & Wrong budget; candidate omission & Independent comparison and constraint checks \\
\textbf{S5 -- Action and Control} & Commands, timing, limits, authority, recovery & Delay, saturation, unsafe translation, failed handoff & Contact; late intervention & Closed-loop outcome, latency, recovery \\
\textbf{S6 -- Feedback and Update} & Logs, updates, checkpoints, rollback, traceability & Contamination, forgetting, backdoor, drift & Update regression; failed rollback & Lineage, regression, persistence, rollback \\
\end{longtable}

These properties are related but not substitutable. Input robustness does not establish the safety of imagined futures. A calibrated predictor does not correct a misspecified cost. A planner that reports no constraint violations in the model may still be unsafe if the model omits the relevant event. Conversely, a runtime shield may prevent observed harm while leaving an upstream defect undiagnosed. A credible claim must specify both the protected object and the point at which the evidence is collected.

Two properties cut across the entire loop. \textbf{Decision-critical fidelity} is the preservation of environmental distinctions and uncertainties that can change plan admissibility, ranking, or safety margin. \textbf{Recoverability} is the existence of a feasible and timely path from a degraded state to an acceptable operating state. Fidelity reduces the probability of unsafe selection; recoverability limits consequence when prediction and prevention fail.

\subsection{From Deviation to Harm: Causal Propagation}\label{from-deviation-to-harm-causal-propagation}

Let \(z_t\) denote the internal state, \(p_\theta\) the learned predictive distribution, and \(\hat{J}_c(\alpha\mid z_t)\) the estimated safety cost of a candidate plan \(\alpha\). A risk source \(r\), whether accidental or adversarial, can be represented by the propagation chain

\[
\begin{aligned}
r &\rightarrow \Delta z_t \rightarrow \Delta p_\theta\\
&\rightarrow \Delta \hat{J}_c \rightarrow \Delta \operatorname{rank}(\alpha)\\
&\rightarrow \Delta a_t \rightarrow H \rightarrow \Delta \mathcal{D}_{t+1},
\end{aligned}
\]

where \(H\) is an external harm or constraint violation and \(\Delta \mathcal{D}_{t+1}\) is the resulting change in future observations, memory, or training data. Not every chain contains every term. A poisoned cost model may act directly on plan ranking, while an execution-channel attack may bypass the learned dynamics. The chain is a diagnostic scaffold for identifying the first corrupted object, the intermediate decision effect, the realized consequence, and any persistence through feedback.

This decomposition motivates \textbf{decision-critical error} as a bridge between predictive and behavioral evaluation. For a plan set \(\mathcal{A}_t\), an error is decision critical if it changes at least one safety-relevant decision relation, such as the feasibility of a plan, the ordering of the safest candidates, the estimated risk budget, the selected action, or the minimum predicted safety margin. The definition is conditional on the task, state, candidate distribution, planner, and constraints. It therefore avoids treating all pixel or latent errors as equally important.

For a concrete coding test, let \(G(M,z_t,\mathcal{A}_t)\) return the vector of safety-relevant relations induced by model \(M\). Comparing a reference model \(M\) with a perturbed or updated model \(M'\), we define

\[
\operatorname{DCE}(M,M')=\mathbf{1}\!\left[G(M,z_t,\mathcal{A}_t)\neq G(M',z_t,\mathcal{A}_t)\right].
\]

The indicator is not a universal safety metric: it must be paired with the relation being changed, the planner, the horizon, and the realized outcome. It is useful because it turns an otherwise vague claim of ``prediction degradation'' into a test of whether a safety-relevant decision actually changes.

The chain is a trace, not a claim that failures are strictly linear. In the coding scheme, each deviation is assigned a \textbf{first corrupted object}, a set of downstream affected objects, and any feedback edge that makes the deviation persistent. We distinguish at least five mechanisms: (i) state or representation error, (ii) dynamics or uncertainty error, (iii) objective/specification error, (iv) planner exploitation of unsupported predictions, and (v) execution or update error. This distinction prevents a runtime collision, for example, from being attributed automatically to the predictive model when the model was adequate but the safety cost or action decoder was incorrect.

Three sensitivity tests follow. First, \textbf{state sensitivity} asks which omissions or perturbations to the internal state change the decision. Second, \textbf{rollout sensitivity} asks how uncertainty, horizon, and multi-step error change candidate evaluation. Third, \textbf{planner sensitivity} asks whether optimization actively selects trajectories with high epistemic uncertainty or known model disagreement. These tests connect model diagnostics to behavior without assuming that a better average prediction metric automatically yields a safer controller.

The chain also clarifies intervention timing. Preventive controls act before or at the first deviation, such as data curation and robust state estimation. Detection controls identify disagreement, uncertainty, anomalous state transitions, or constraint risk during inference. Mitigation controls reduce exposure by shortening rollouts, applying conservative objectives, or restricting action. Recovery controls move the system to a fallback state and restore a trusted version. Assurance controls establish that these mechanisms work under declared assumptions. Defense in depth is required because no single stage can validate itself against all of its own blind spots.

\subsection{Threat Conditions and Operating Assumptions}\label{threat-conditions-and-operating-assumptions}

An attack claim is meaningful only under an explicit threat model. We characterize adversaries along six dimensions.

\begin{enumerate}
\def\labelenumi{\arabic{enumi}.}
\tightlist
\item
  \textbf{Objective:} integrity violation, targeted unsafe behavior, hazard concealment, plan-ranking manipulation, availability degradation, information extraction, or persistent compromise.
\item
  \textbf{Lifecycle phase:} data collection, pretraining, fine-tuning, deployment inference, action execution, or online update.
\item
  \textbf{Access surface:} training examples, observations, memory, parameters, model queries, planning interface, action channel, logs, or feedback.
\item
  \textbf{Knowledge:} white-box, gray-box, or black-box knowledge of the model, planner, defense, deployment version, and environment.
\item
  \textbf{Capability and budget:} read or write permission, poisoning fraction, perturbation magnitude, query count, physical realizability, temporal synchronization, duration, and collusion across modalities.
\item
  \textbf{Persistence and adaptivity:} one-step or sequential influence, fixed or adaptive strategy, and survival across memory refresh, fine-tuning, checkpoint replacement, or rollback.
\end{enumerate}

These dimensions extend established adversarial-machine-learning distinctions \citep{biggio2018wildpatterns} to sequential prediction and control. Backdoor work on deep reinforcement-learning agents shows why the lifecycle and persistence dimensions matter: a policy can appear benign on ordinary trajectories yet change behavior when a trigger is present \citep{kiourti2019trojdrl}. Attack success must be reported at the deepest demonstrated causal level: internal model change, rollout change, plan change, action change, realized safety consequence, or persistence after update. A change in predicted pixels is not by itself a successful control attack. Conversely, a small input perturbation that reliably reverses a plan ranking and induces a constraint violation should not be evaluated only by perceptual distance.

For adaptive evaluations, the defense configuration, model version, query budget, and attacker observations must be held out explicitly. A study that tests only a fixed perturbation against a fixed defense is coded as a non-adaptive attack; it cannot support a claim of robustness to an attacker who knows the defense or can iterate over time. Sequential attacks are additionally coded for dwell time, recovery after intervention, and whether the effect survives memory refresh, checkpoint replacement, or rollback.

Non-adversarial evaluation requires a parallel description of operating conditions without inventing an attacker. The protocol should state the training and test distributions, environment variability, disturbance and sensor models, observability, model version, update status, planning horizon, action frequency, reset assumptions, and consequence severity. Long-tail or out-of-distribution operation is non-adversarial when it arises from the environment rather than deliberate manipulation. The same functional interfaces can thus index failures and attacks, while their causal premises remain separate.

This distinction has practical consequences for defense evaluation. Adversarial defenses must consider adaptive attackers, transfer, stealth, and defense knowledge. Safeguards for non-adversarial failures must consider calibration, coverage, worst-case or tail behavior, monitoring, and graceful degradation. A mechanism may address both, but evidence against random corruption cannot be presented as evidence against an adaptive adversary.

\subsection{Evidence Directness, Outcome Levels, and Claim Discipline}\label{evidence-directness-outcome-levels-and-claim-discipline}

World-model safety claims require outcome evidence at four levels.

\textbf{Model-level evidence} concerns the correctness and uncertainty of state and future prediction. Metrics may include hazard-variable recall, state identifiability, multi-step error, negative log-likelihood, calibration error, out-of-distribution detection, physical consistency, counterfactual consistency, and event coverage. Model-level evidence localizes a defect but does not establish its decision consequence.

\textbf{Planning-level evidence} concerns the use of the model by an optimizer or learner. Metrics include candidate-ranking stability, regret under the true system, predicted and realized constraint violation, risk-budget error, feasibility, sensitivity to uncertainty, and the frequency with which planning enters unsupported regions. This level is the principal bridge between representation and action.

\textbf{Behavior-level evidence} concerns executed outcomes. It includes task completion, collision or hazardous contact, minimum safety margin, rule violation, emergency intervention, near misses, and successful recovery. Behavior-level evidence is closer to harm, but it remains conditional on scenario coverage, controller design, reset policy, and test duration.

\textbf{Operational-level evidence} concerns sustained and changing deployment. It includes detection latency, intervention frequency, recovery time, update-induced regression, rollback success, attack persistence, reliability across versions, and performance under degraded sensing or communication. This level tests whether safety can be maintained rather than achieved once.

The levels are complementary and do not automatically imply one another. Improved calibration supports but does not prove safer planning; uncertainty estimates can themselves degrade under dataset shift \citep{ovadia2019uncertainty}. Absence of collision in a finite benchmark does not prove that the model represents every relevant hazard. A formal guarantee applies only under its state, dynamics, uncertainty, constraint, and disturbance assumptions \citep{berkenkamp2017safe, hewing2020learningbased}. A runtime shield can establish bounded behavior under its intervention authority, but it does not certify an upstream learned model. Every strong claim should therefore state the object, operating domain, assumptions, horizon, adversary if any, outcome level, and remaining failure modes.

Accordingly, each study is summarized using the following claim template: \textbf{under} {[}operating assumptions and model/deployment version{]}, \textbf{the method changes} {[}protected object or decision relation{]}, \textbf{as measured by} {[}metric and test distribution{]}, \textbf{with the deepest demonstrated consequence at} {[}model, planning, behavior, or operational level{]}. If any field is missing, the claim is marked incomplete rather than upgraded by interpretation.

Separately, we classify the \textbf{directness} of the literature evidence. Directness is not a quality ranking and must not be confused with outcome level: a direct E1 study may report only model-level evidence, while an interface E2 study may report stronger closed-loop behavior evidence.

\begin{itemize}
\tightlist
\item
  \textbf{E1: Direct world-model evidence.} The evaluated object is a learned environmental predictor, its state representation, imagined rollout, or its explicitly studied contribution to learning, planning, or monitoring. This includes learned dynamics in model-based reinforcement learning and explicit occupancy world models; neither a particular architecture nor a physical-safety result is required for E1.
\item
  \textbf{E2: Interface evidence.} The evaluated object is a specific neighboring observation, forecasting, planning, control, or learning interface that can transmit world-model risk, without directly evaluating the world-model dependency in question. A standalone trajectory predictor or a predictive filter using assumed dynamics can supply E2 evidence. A study is not E2 merely because it calls its learned world model a dynamics model.
\item
  \textbf{E3: System evidence.} The evaluated object is the behavior or assurance of an integrated agent or controller, without isolating the world-model dependency relevant to the claim. E3 may come from simulation or hardware; hardware deployment alone does not change E1 or E2 into E3.
\item
  \textbf{E4: Background evidence.} A source supplies terminology, a general mechanism, or secondary synthesis rather than a direct test of the world-model or interface claim. Secondary syntheses and general governance guidance are E4 when used in this role.
\end{itemize}

The coding unit is a claim supported by a particular analysis or experiment, not a permanent label for an entire paper. A paper can supply E2 evidence for a controller interface and E3 evidence for a separately discussed system demonstration, but each assignment must identify the corresponding result. Repeated use of the same result retains its classification across sections. Where roles overlap, use E1 for an explicitly evaluated learned predictive dependency, E2 for an isolated adjacent interface, E3 for an integrated-system result, and E4 for background use. These are evidence categories, not a quality ranking, and their numerical labels are not scores.

Outcome level is recorded separately. Physical task execution can be behavior-level evidence; it becomes operational-level evidence only for an evaluated monitoring, intervention, maintenance, persistence, or update process. A controlled runtime anomaly-detection experiment can supply operational-level detection evidence without establishing successful recovery or long-term reliability. Likewise, offline evaluation of candidate plans is planning-level evidence even when no closed-loop trajectory is executed. Guarantee type (empirical, probabilistic, robust, or formal) and publication status remain additional attributes.

Evidence directness and outcome level are intentionally kept as prose categories rather than another large table. Directness identifies what dependency was isolated: \textbf{E1} direct world-model evidence, \textbf{E2} adjacent interface evidence, \textbf{E3} integrated-system evidence, and \textbf{E4} background or conceptual evidence. Outcome level identifies how far the evaluation reaches: \textbf{model}, \textbf{planning}, \textbf{behavior}, or \textbf{operational}. The two axes are reported together but neither is a safety score; a direct E1 study can remain model-level, while an E2 or E3 study can reach closed-loop behavior without isolating the world-model cause.

\subsection{Unified Coding Schema and Cross-Sectional Use}\label{unified-coding-schema-and-cross-sectional-use}

The six-stage label is the primary index used to organize the later synthesis. Each study is coded using the tuple

\[
\begin{aligned}
&\text{model family and use mode} \times \text{primary loop stage}\\
&\times \text{risk mechanism} \times \text{protected object}\\
&\times \text{defense function} \times \text{evaluation level}\\
&\times \text{domain} \times \text{evidence directness}.
\end{aligned}
\]

\textbf{Model family and use mode} use the categories introduced in Section 2. \textbf{Primary loop stage} records the first supported point of deviation or intervention, while the object analyzed and downstream affected stages are recorded separately. A visual attack can enter at S1 and be discussed under S4 because its ranking effect is the object of analysis; this does not establish direct write access to the planner. When the evidence cannot locate a unique first deviation, attribution remains unresolved or set-valued. \textbf{Risk mechanism} distinguishes non-adversarial failure from attack. \textbf{Protected object} uses the vocabulary in Table~\ref{tab:six-stage-map}, while \textbf{defense function} distinguishes prevention, detection, mitigation, isolation, recovery, and assurance renewal after change. \textbf{Evaluation level} and \textbf{evidence directness} use the definitions in Section 3.6. Domain, operating assumptions, and publication status are recorded as separate metadata rather than folded into the stage label.

The schema allows a study to be discussed under one primary stage while preserving cross-stage propagation. It also connects attacks, defenses, and metrics: an S1 input defense evaluated only through reconstruction quality cannot automatically support a behavior-level claim at S5. Aggregating the coded records by stage exposes whether a part of the loop contains many attack demonstrations but few defenses, or model-level metrics without planning, behavior, or operational evidence.

An illustrative coding record is: \textbf{OccWorld} \citep{zheng2024occworld} -\textgreater{} family: object-centric/spatial; use mode: model-mediated prediction and planning; analyzed stage: S3 Dynamics and Imagination, with downstream S4 planning; protected object: occupancy and reachable free space; candidate risk mechanisms: occlusion, tracking, or discretization error, without assuming a diagnosed root cause for every case; defense function: not assumed unless evaluated; outcome level: model-level occupancy prediction and planning-level offline ego-trajectory evaluation; directness: E1 because the learned occupancy world model is the evaluated object. These results do not by themselves establish executed closed-loop driving safety.

Domain is coded as an operational constraint rather than a decorative label. For each domain, the later synthesis records consequence severity, action and sensing latency, observability, reset availability, human proximity, and recovery authority. Thus the same plan-ranking error can be a low-cost benchmark failure in one domain but an unrecoverable safety violation in another. Cross-domain comparisons are made only after these conditions are reported; raw task-success percentages are not treated as comparable safety evidence.

Sections 4 and 5 use S1--S6 to trace non-adversarial failures and attacks from entry point to consequence. Section 6 maps preventive, runtime, and recovery defenses to the same stages without assuming a one-to-one correspondence between one attack and one defense. Section 7 evaluates the resulting claims through the four evidence levels, and Section 8 examines how domain conditions alter the meaning of each stage and transition. The repeated use of S1--S6 makes the framework, rather than an architectural family or application domain, the common structure of this work.

The framework also makes its limitations explicit. It cannot infer real-world safety from model-level metrics alone, cannot resolve an underspecified safety objective, and cannot make evidence from one domain automatically transferable to another. Claims are therefore reported with their operating assumptions, model and deployment versions, horizon, intervention authority, and deepest demonstrated outcome. These qualifiers are part of the result rather than editorial footnotes.

\FloatBarrier
\section{Non-Adversarial Failures Across the World-Model Lifecycle}\label{non-adversarial-failures-across-the-world-model-lifecycle}

\subsection{Scope, Attribution, and Organization}\label{scope-attribution-and-organization}

A world model can influence a decision through a chain of observations, internal state, predicted consequences, plan evaluation, executed action, and subsequent feedback. Each link can fail without an attacker. A rare condition may be absent from experience; a latent state may discard an occluded hazard; a rollout may predict the wrong consequence of an action; or a controller may execute a sound plan too late. These mechanisms differ even when they end in the same constraint violation. This section uses the six stages defined in Section 3 to locate the first unreliable safety-relevant object: S1 Data and Observations, S2 State and Memory, S3 Dynamics and Imagination, S4 Evaluation and Planning, S5 Action and Control, and S6 Feedback and Update.

Our attribution rule assigns a failure to the earliest stage at which a safety-relevant object becomes corrupted. Later stages through which the error travels are recorded as propagation rather than counted again as independent root causes. Insufficient coverage of an action in training data is an S1 failure if it precedes an incorrect action consequence in S3, an optimistic ranking in S4, and a constraint violation in S5. If the predicted consequence is accurate but the planner ignores a known risk, the first corrupted object is the plan ranking in S4. If the plan is sound but the actuator response differs from the planned action, the primary stage is S5. The rule disciplines analysis; it does not imply that every reported system failure has a uniquely identifiable cause. A final collision or low task return cannot by itself establish which upstream object first failed.

The evidence-directness and outcome-level conventions defined in Section 3.6 apply throughout this section. We retain the result-level coding there rather than redefining E1--E4 or the model, planning, behavior, and operational levels here. When a study cannot identify a unique first corrupted object, the primary-stage attribution remains unresolved; the mechanism discussed below is then a conditional explanation rather than an established root cause.

The safety interpretation must respect both axes. Task return and visual quality are not substitutes for constraint satisfaction; a simulated cost violation is not an observed physical injury; and identifying an anomaly does not establish timely recovery. We therefore describe what each body of work actually observes and where a causal link remains conditional. The sections analyze failure formation and propagation without cataloguing defense designs or benchmark protocols.

\begin{figure}[!htbp]
\centering
\includegraphics[width=\linewidth,height=0.36\textheight,keepaspectratio]{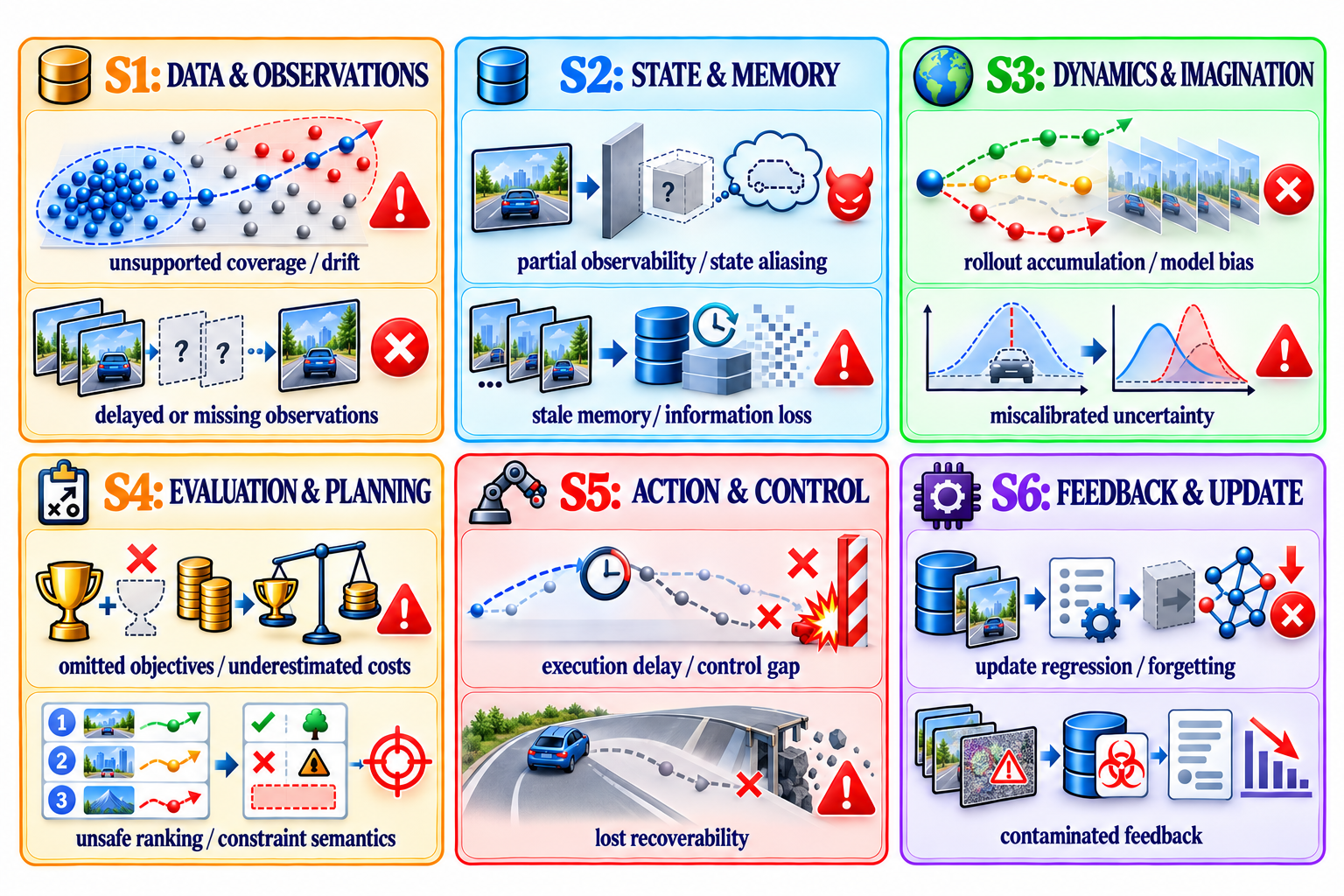}
\caption{Non-adversarial failure mechanisms across the six-stage lifecycle.}
\Description{Six panels group representative non-adversarial failure mechanisms by lifecycle stage and show possible downstream propagation between stages.}
  \label{fig:failure-lifecycle}
\end{figure}

\subsection{S1 --- Data and Observations: Coverage, Distribution Shift, and Timing}\label{s1-data-and-observations-coverage-distribution-shift-and-timing}

\subsubsection{Missing Support for Safety-Relevant States and Actions}\label{missing-support-for-safety-relevant-states-and-actions}

The first coverage question is whether the experience used to fit a world model contains the state--action pairs on which future decisions will rely. A dataset may contain many observations of a scene yet few outcomes for a particular action in that scene. The gap concerns action consequences, not simply visual diversity or dataset size. When a planner evaluates an unsupported action, S3 must extrapolate its consequences before S4 ranks the resulting candidate. The risk is most consequential when the unsupported transition concerns a safety-relevant boundary, although ordinary task metrics do not identify such boundaries automatically.

Studies of robust world-model training and offline model-based reinforcement learning approach this gap from different directions. WAKER reports variation in model error across training environments; MOPO and MOReL analyze the mismatch between a fixed collection policy and actions selected by a learned policy \citep{rigter2024waker, yu2020mopo, kidambi2020morel}. Together they support the claim that training coverage is conditional on both encountered environments and selected actions. Their E1 evidence covers model- and behavior-level results in the studied tasks. It does not show that a particular dataset covers rare hazards or establish operational accident frequencies. Policy-induced shift also differs from natural environmental change: the former is produced by optimization, whereas the latter changes the conditions under which observations or transitions are generated.

\subsubsection{Distribution Drift and Changing Physical Conditions}\label{distribution-drift-and-changing-physical-conditions}

The environment can leave the training support without a policy deliberately selecting unusual actions. Background appearance, physical properties, and transition dynamics can change after data collection. A model fitted under one regime then receives observations or action consequences from another. This is an S1 coverage failure when the changed conditions were absent from the data entering the model; errors that arise despite adequate coverage belong to S2 or S3. A good fit to historical transitions provides no direct evidence about the changed regime.

WAKER examines variation across environments, while Augmented World Models tests policies under changed dynamics after learning from one offline environment \citep{rigter2024waker, ball2021augmented}. RePo and Policy-shaped Prediction highlight a complementary visual problem: predictable or changing background content can consume a visual world model's representational capacity or disrupt performance even when it is irrelevant to the task \citep{zhu2023repo, hutson2024policy}. These E1 studies show model- and behavior-level sensitivity to environmental variation. They do not jointly measure hazard severity, and a visual distractor that reduces reward is not automatically a safety violation. They do reject the inference that accurate reconstruction alone establishes robust coverage of safety-relevant conditions.

\subsubsection{Incomplete and Temporally Misaligned Observations}\label{incomplete-and-temporally-misaligned-observations}

At deployment, the observation delivered to the model may be missing, noisy, or older than the current physical state. A stale sensor packet is an S1 error when its timestamp or availability is wrong before state estimation begins; a correct packet misinterpreted by the belief or memory mechanism is an S2 error. With multiple sensors, inconsistent timing can also make individually plausible measurements jointly describe no single state. The immediate corrupted object is the observation stream. Propagation occurs when S2 treats it as current, S3 projects it forward, and S4 evaluates actions from the wrong starting point.

\citet{chen2023impaired} formally study acting with delayed or missing state observations, showing why impaired availability changes the information on which a policy can condition. \citet{holt2023laplace} study continuous-time control with irregular observations and unknown action delay in a learned-dynamics setting. The first supplies E2 interface evidence and the second supplies E1 model-based evidence; both report behavior-level outcomes in their respective settings. Neither isolates a multi-sensor world-model failure or demonstrates physical harm caused by desynchronization. The claim concerns the broken temporal assumption at the observation interface, not a measured rate of real-world accidents.

\subsection{S2 --- State and Memory: Observability, Representation, and Memory Staleness}\label{s2-state-and-memory-observability-representation-and-memory-staleness}

\subsubsection{Partial Observability and State Aliasing}\label{partial-observability-and-state-aliasing}

An observation may be accurate yet insufficient to identify the current state. Under partial observability, distinct physical situations can produce the same available input. State aliasing occurs when the internal state also fails to distinguish those situations, whether because the observation is insufficient, the model compresses away a critical feature, or history has not been integrated correctly. The first corrupted object is the state estimate or belief. If aliased states call for different actions to satisfy a constraint, downstream predictions can be internally consistent while guiding the wrong choice.

Safe pixel-based latent-state learning and PIGDreamer both study decisions under partial observability while separating task outcomes from safety costs \citep{hogewind2023pixels, huang2025pigdreamer}. The lambda-discrepancy analysis gives complementary E2 evidence that a non-Markovian representation can be detected through disagreement between value estimates that depend differently on history \citep{allen2024lambda}. These studies support model- and behavior-level consequences of missing state information in the investigated environments. They do not show that every visual ambiguity has a safety consequence or that privileged training information remains available at deployment. The narrower requirement is that a state used for prediction retain distinctions on which safety-relevant action consequences depend.

\subsubsection{Safety-Relevant Information Loss Through Compression}\label{safety-relevant-information-loss-through-compression}

A latent state cannot preserve every detail of an observation history. Compression becomes a safety failure when it discards information needed to predict a hazard, even if it retains enough for average reward or image reconstruction. The opposite allocation error is possible: a model can spend capacity on predictable but task-irrelevant details, leaving less capacity for decision-relevant dynamics. This differs from S1 coverage. The data may contain the signal, but the mapping from observation history to internal state fails to carry it forward.

RePo reports susceptibility of visual model-based agents to spurious variation and limits of learned invariance under significant distribution shift; Policy-shaped Prediction reports the persistent effect of highly predictable visual distractions on leading model-based agents \citep{zhu2023repo, hutson2024policy}. These E1 observations include behavior-level task outcomes and support the mechanism that representation objectives and control needs can diverge. They do not establish which latent features encode a physical hazard in an unconstrained deployment. In particular, removing a feature because it is irrelevant to reward is not evidence that it is irrelevant to safety cost. Reward and safety can impose different representation requirements.

For systems that expose a language-based description of state, factual
accuracy and faithfulness to the supplied context are distinct
requirements. SPACE jointly addresses these two forms of hallucination
by editing shared activation subspaces in LLMs \citep{wang2025space}.
This provides an E4 mechanism-level analogy for auditing semantic state
reports: fluent or factually plausible text can still omit the
observation that matters for a decision. It does not establish the
adequacy of a learned physical state.

\subsubsection{Long-Horizon Memory and Stale Beliefs}\label{long-horizon-memory-and-stale-beliefs}

History helps distinguish states that look identical at one instant, but it introduces another failure route. A remembered object, map entry, or latent context can become stale when the world changes without a corrective observation. Conversely, limited memory can forget an earlier observation before its consequence becomes relevant. The first error is in S2 if incoming observations are valid but the stored belief no longer represents the present. S3 may then make a faithful prediction from the wrong state, and S4 may rank plans using a hazard location that is no longer current.

Comparisons of recurrent, transformer, and state-space world-model backbones explicitly test long-term imagination, recall, and memory-based reasoning \citep{deng2023backbones}. PIGDreamer and the lambda-discrepancy work show how incomplete observation histories alter decision-relevant state estimates \citep{huang2025pigdreamer, allen2024lambda}. The world-model studies provide E1 evidence, while the general representation analysis provides E2 evidence; together, they report model- or behavior-level results. They do not establish persistent memory drift in deployed world-model systems or a measured link from that drift to operational harm. The studies justify distinguishing a memory failure from a new dynamics error, but do not close the full S2-to-S5 causal chain.

\subsection{S3 --- Dynamics and Imagination: Rollout Error and Uncertainty}\label{s3-dynamics-and-imagination-rollout-error-and-uncertainty}

\subsubsection{Local Model Bias and Rollout Accumulation}\label{local-model-bias-and-rollout-accumulation}

A dynamics error first appears when a model predicts the wrong next state, observation, reward, or cost for a given state and action. During an imagined rollout, the next predicted state becomes the input to another prediction. Local errors can therefore alter later predictions, particularly when the imagined trajectory moves beyond the data that trained the model. This is error accumulation inside the model. It differs from closed-loop amplification, in which an executed action changes the real state and therefore the next real observation.

MBPO makes the dependence on rollout horizon explicit, and MACURA challenges the assumption that model accuracy is uniform across state space \citep{janner2019mbpo, frauenknecht2024macura}. Talvitie examines a related failure of reward evaluation on erroneous states generated by a misspecified model \citep{talvitie2018reward}. These E1 studies support model- and behavior-level consequences of compounding or locally concentrated model error. Their performance results do not measure distant hazards directly. A short useful rollout also does not imply that a long-horizon safety condition is satisfied: the unmodeled tail still influences any decision whose consequence lies beyond the reliable prediction horizon.

\subsubsection{Plausible Predictions with Incorrect Action Consequences}\label{plausible-predictions-with-incorrect-action-consequences}

The safety-relevant question is often counterfactual: would a different candidate action from the same state have produced a different outcome? A model may fit the average next observation or produce a coherent future while failing to preserve the ordering of candidate actions. In that case, a planner can prefer the wrong action even though conventional prediction metrics look acceptable. The first corrupted object is the action-conditioned transition or cost prediction in S3; the incorrect choice in S4 is its propagated effect.

Objective Mismatch directly finds that one-step dynamics likelihood need not correlate with downstream control performance \citep{lambert2020mismatch}. Transition Occupancy Matching studies why a model fitted indiscriminately to past transitions need not match the transitions encountered by the current policy \citep{ma2023tom}. Talvitie's reward-model analysis further shows that values assigned to model-generated states affect the usefulness of learned dynamics \citep{talvitie2018reward}. These E1 results provide model- or behavior-level evidence and establish a decision-relevance problem, not a general physical-safety rate. Prediction accuracy and decision utility are separate properties. Evidence that a model reproduces appearance alone is weaker still unless action responsiveness and the resulting choices are tested.

\subsubsection{Uncertainty Miscalibration and Unsupported Confidence}\label{uncertainty-miscalibration-and-unsupported-confidence}

Uncertainty itself is not a failure. A stochastic environment can remain uncertain under a well-specified model. Failure occurs when reported confidence does not match the frequency or size of prediction errors, especially in state--action regions poorly represented by data. If an overconfident model supplies narrow forecasts there, S4 can rank a plan as reliable when its consequences are not. This is an S3 failure when the uncertainty estimate is wrong; an accurate warning that the planner ignores is an S4 failure.

Calibration work in model-based reinforcement learning shows that predictive uncertainty can be inaccurate and affect planning, while COMBO reports settings in which explicit uncertainty estimates in offline model-based learning are unreliable enough to impair performance \citep{malik2019calibrated, yu2021combo}. MOPO and MOReL locate the problem at the edge of offline data support \citep{yu2020mopo, kidambi2020morel}. The direct E1 evidence covers model- and behavior-level results. It does not establish calibration specifically on rare hazards, and average calibration does not prove that the subset of plans selected for execution is calibrated. That selected subset matters because S4 repeatedly exposes the system to predictions it finds most attractive.

\subsection{S4 --- Evaluation and Planning: Specifications, Constraints, and Plan Selection}\label{s4-evaluation-and-planning-specifications-constraints-and-plan-selection}

\subsubsection{Omitted Objectives and Underestimated Costs}\label{omitted-objectives-and-underestimated-costs}

Planning can fail even when the world model predicts physical consequences accurately. A specification error occurs when a hazard is absent from the reward, cost, or constraint used to evaluate plans. The first corrupted object is then the evaluation criterion in S4. Cost underestimation is different: the hazard is specified, but the predicted state or cost makes it appear smaller than it is. That error may originate in S3, with its unsafe ranking propagated through S4. Keeping these cases separate prevents an inaccurate world model from being blamed for a missing safety objective, or a complete objective from being credited for inaccurate predictions.

CAP studies how errors in learned dynamics can make real cumulative costs differ from costs estimated in model-based safe learning \citep{ma2022cap}. SafeDreamer evaluates reward and safety cost within a world-model setting, and CASE reports that constrained model-based policies can still incur violations during learning and exploration \citep{huang2024safedreamer, agha2025case}. These E1 studies report planning- or behavior-level outcomes under their specified costs. They do not demonstrate that those costs exhaust all hazards in deployment. Their shared evidence supports the narrower conclusion that satisfying an internally represented cost condition does not prove that the model predicted the real cost correctly or that the specification expressed every relevant harm.

\subsubsection{Constraint Semantics and Risk Aggregation}\label{constraint-semantics-and-risk-aggregation}

The form of a constraint matters independently of its numerical estimate. An expected cumulative-cost budget restricts an average over trajectories. A chance constraint concerns the probability of an event, while a per-step or per-trajectory rule concerns what may happen in an individual execution. These statements are not interchangeable. A planner can satisfy an expectation while permitting a small set of severe violations; conversely, a risk-sensitive rule may reject a plan with acceptable mean cost. When the wrong aggregation is chosen for the intended safety requirement, the first corrupted object is the criterion used to evaluate candidates in S4.

Model-based constrained learning, chance-constrained model predictive control, and analyses of long-term average constraints instantiate different formal safety quantities \citep{ma2022cap, pfrommer2022chance, agarwal2022average}. They provide E1 or E2 planning-level evidence about what each formal statement establishes under its assumptions. The distinction is logical, not an empirical claim that one formulation is universally safer. A theorem about expected or chance-constrained behavior cannot be read as a guarantee of zero violations on every realized trajectory, and a simulation of constraint costs cannot identify omitted real-world hazards.

\subsubsection{Model Exploitation and Ignored Uncertainty}\label{model-exploitation-and-ignored-uncertainty}

Optimization can amplify a model error because it searches over candidate actions and preferentially selects those with the most favorable predicted outcomes. This selection effect differs from merely encountering an unfamiliar state. If a candidate's optimistic prediction is caused by S3 error, exploitation describes its propagation through S4. If the model already reports that the candidate is unreliable but the planner treats its forecast as certain, the earliest new error is in S4's use of uncertainty. The mechanisms can coexist, but they call for different attribution.

MOPO, MOReL, and COMBO all analyze the difficulty of planning beyond offline data support; COMBO additionally reports cases in which explicit uncertainty estimates do not provide reliable guidance \citep{yu2020mopo, kidambi2020morel, yu2021combo}. CAP connects optimism about learned safety costs with constraint feasibility \citep{ma2022cap}. This E1 evidence covers planning- and behavior-level outcomes in the studied environments. It supports selection-induced exposure to model error, but does not isolate the world model as the cause of every later system failure. A conservative estimate of task return does not repair an omitted safety objective, and uncertainty cannot guide choices when the evaluation rule ignores it.

\subsection{S5 --- Action and Control: Execution, Timing, and Recovery}\label{s5-action-and-control-execution-timing-and-recovery}

\subsubsection{Plan-to-Execution Gaps and Reality Mismatch}\label{plan-to-execution-gaps-and-reality-mismatch}

A plan may be valid for the predicted transition and still fail when the commanded action and physical action differ. Unmodeled actuation, contact, or plant dynamics can change the trajectory after planning; a policy transferred from simulation may also encounter dynamics outside those represented during training. When the model's prediction was correct for the action it assumed, but the action delivered to the environment differs, the first corrupted object is the executed action or its timing in S5. When the plant transition differs even for the same executed action, the primary error is instead in S3's dynamics model. This distinction requires action and state logs, not merely a final task result.

Sim-to-real studies document behavior changes associated with discrepancies between simulated and real robot trajectories, while Augmented World Models examines changing physical dynamics after offline training \citep{golemo2018neural, ball2021augmented}. TRANSIC uses human corrections to expose sim-to-real gaps in a complete robotic policy, supplying E3 evidence from a complete system rather than a diagnosis of a particular world model \citep{jiang2025transic}. Together these studies report behavior-level outcomes, including physical-robot experiments, but their attribution differs. A transfer failure cannot automatically be classified as a world-model prediction failure, and better transfer performance does not establish satisfaction of an unmeasured safety constraint.

\subsubsection{Delay, Saturation, and Loss of Recoverability}\label{delay-saturation-and-loss-of-recoverability}

Control latency changes where a robot will be when a chosen action takes effect. An actuator limit can similarly make the applied command differ from the planned one. If either deviation consumes the margin assumed by a plan, a state predicted to remain recoverable may leave the set from which a specified backup action can return safely. The immediate S5 error is the mismatch between planned and executable control. Its consequence depends on the current state, delay or disturbance, and recovery condition; delay alone is not evidence of a violation.

Real-robot model-based locomotion explicitly addresses planning latency, and Neural Laplace Control studies delayed action effects and irregular observations in a learned-dynamics setting \citep{yang2020legged, holt2023laplace}. \citet{berkenkamp2017safe} analyze stability within a conditionally verified region; Robust Model Predictive Shielding analyzes recovery under its specified nominal dynamics, disturbance distribution, controller, and finite horizon \citep{li2020shielding}. These E1 and E2 studies report planning- or behavior-level outcomes. They support the importance of timing and model assumptions for recoverability, not a general guarantee that any delayed system can be returned to safety. Once execution leaves the verified region, the original conditional claim no longer applies; that does not prove recovery is impossible by every controller.

\subsubsection{Switching at the Wrong State or Time}\label{switching-at-the-wrong-state-or-time}

A controller switch is an S5 failure if the backup action actually applied is inconsistent with the state, controller history, or timing for which recovery was evaluated. The condition is stricter than detecting that a prediction was wrong. A monitor may raise a correct alarm after the trajectory has crossed a boundary beyond which its backup plan was verified, or a switch may resume a controller state different from the one used in the planning check. In either case, the first corrupted object is the executable recovery decision.

Robust Model Predictive Shielding distinguishes recoverability under specific backup-controller conditions \citep{li2020shielding}. World-model anomaly detection compares expected and observed behavior in simulation and in a controlled real-quadcopter experiment \citep{domberg2025anomaly}. The first supplies E2 evidence for conditional recovery; the second supplies E1 operational-level evidence restricted to runtime anomaly detection. The latter classification concerns the monitoring process, not the mere use of hardware. These results do not establish the complete chain from detection through a timely switch to prevention of physical harm. Detection and successful recovery are separate outcomes.

\subsection{S6 --- Feedback and Update: Feedback Attribution, Contamination, and Regression}\label{s6-feedback-and-update-feedback-attribution-contamination-and-regression}

\subsubsection{Prediction--Observation Disagreement and Attribution}\label{predictionobservation-disagreement-and-attribution}

After execution, a world-model system may compare what it predicted with what occurred. A discrepancy can be informative, but its meaning depends on whether the recorded action, observation time, and intervention state are correct. A residual between prediction and observation can arise from a dynamics change, stale observation, unmodeled disturbance, or action that was never executed as proposed. Treating every residual as a model fault would corrupt the diagnosis that guides later updates. The first S6 failure is in the feedback record or its attribution when earlier stages were correctly recorded but the update process mislabels their outcome.

\citet{domberg2025anomaly} show that world-model prediction discrepancies can identify simulated dynamics changes and unexpected forces on a quadrotor. This E1 operational-level evidence concerns runtime anomaly identification, not longitudinal reliability or successful recovery. The inference from a residual to a unique root cause remains conditional. The S1-to-S6 attribution rule therefore applies again during feedback: later disagreement requires examination of observation timing, executed control, and dynamics before the data are assigned to a particular model failure. The cited detection result does not establish that all anomalies were identified early enough for recovery.

\subsubsection{Intervention-Contaminated Transition Records}\label{intervention-contaminated-transition-records}

Safety interventions create a specific ambiguity in learning from experience. A task policy can propose action \(a_p\), an intervention can replace it with executed action \(a_e\), and the environment then produces an outcome from \(a_e\). A physical dynamics model trained on the pair \((a_p, s')\) would learn the consequence of an action that was not applied. A model of the full policy-plus-intervention system may legitimately predict the outcome conditional on \(a_p\), but only if the intervention rule is part of the modeled process. Confusing these targets first corrupts the feedback transition in S6 and can later corrupt the S3 action--consequence relation.

Recovery RL studies the effect of intervention-aware action records on task-policy learning, including an action-relabeling ablation in simulation \citep{thananjeyan2021recovery}. This provides E2 behavior-level evidence for the importance of how intervention data are assigned; it is not a direct demonstration of physical safety degradation from world-model retraining. The distinction between proposed and executed action follows from the causal data-generating process, rather than a claim that the cited study measured a deployed world model with mislabeled transitions. The appropriate record depends on which system the model is intended to predict.

\subsubsection{Update Regression, Forgetting, and Safety Drift}\label{update-regression-forgetting-and-safety-drift}

Feedback changes the system only when it alters data, state, model parameters, or policy. An update may improve performance on recent conditions while degrading a previously supported capability. In a world-model system, forgetting can affect stored transition knowledge, and a changed model can alter which plans appear feasible even if the planning code is unchanged. The first corrupted object belongs to S6 when the update itself removes a previously valid relation or invalidates an earlier safety claim. A new environment regime that merely exposes an unchanged model's limits is instead an S1 or S3 issue.

Continual world-model studies explicitly measure forgetting and transfer under changing tasks, and online world-model planning studies compare how model-update choices affect retention of earlier skills \citep{kessler2023continual, liu2025continual}. Deep SPI analyzes policy improvement using a world model with a term for modeling error \citep{delgrange2026deepspi}. These E1 results report planning- or behavior-level task outcomes. They do not show that a previously safe physical behavior remains safe after every update, because task return is not a safety-cost or recovery-capability measure. The studies examined here support a risk of update regression, while evidence for repeated world-model updates preserving operational safety conditions remains limited in this representative set.

A related feedback problem arises in self-evolving language-agent
communities. The Moltbook study analyzes safety drift under isolated
self-evolution using an information-theoretic formulation and
agent-system observations \citep{wang2026moltbook}. It motivates
checking whether internally recycled experience preserves an independent
safety reference. Its theoretical conclusions depend on that formulation
and remain E4 background here; they do not establish inevitable safety
degradation in a world-model controller with external observations and
interventions.

\subsection{Cross-Stage Propagation and Evidence Gaps}\label{cross-stage-propagation-and-evidence-gaps}

The six-stage organization separates the location of the first error from the path along which it acts. Insufficient action coverage can yield an unsupported S3 transition, which S4 selects because it looks favorable, after which S5 exposes the physical system to the error. An S2 memory lapse can yield an inaccurate starting state for an otherwise valid dynamics model. A sound S4 plan can fail in S5 because the applied action arrives late. After any of these sequences, a mislabeled S6 record can preserve the wrong explanation. These are causal possibilities, not accident reconstructions. A complete attribution requires intermediate observations, state estimates, predicted alternatives, selected and executed actions, and feedback records.

Table~\ref{tab:failure-mechanisms} compares representative works by the failure they expose, the affected stage and condition, and the limit of the evidence. It is a guide to mechanisms rather than an exhaustive bibliography.

\begin{longtable}[]{@{}>{\raggedright\arraybackslash}p{(\linewidth - 45pt) * \real{0.18}}@{\hspace{3pt}}>{\raggedleft\arraybackslash}p{(\linewidth - 45pt) * \real{0.09}}@{\hspace{9pt}}>{\raggedright\arraybackslash}p{(\linewidth - 45pt) * \real{0.12}}@{\hspace{9pt}}>{\raggedright\arraybackslash}p{(\linewidth - 45pt) * \real{0.18}}@{\hspace{9pt}}>{\raggedright\arraybackslash}p{(\linewidth - 45pt) * \real{0.19}}@{\hspace{9pt}}>{\raggedright\arraybackslash}p{(\linewidth - 45pt) * \real{0.24}}@{}}
\caption{Compact comparison of non-adversarial failure mechanisms.}\label{tab:failure-mechanisms}\\
\toprule\noalign{}
\begin{minipage}[b]{\linewidth}\raggedright
Work
\end{minipage} & \begin{minipage}[b]{\linewidth}\raggedleft
Year
\end{minipage} & \begin{minipage}[b]{\linewidth}\raggedright
Stage
\end{minipage} & \begin{minipage}[b]{\linewidth}\raggedright
Failure type
\end{minipage} & \begin{minipage}[b]{\linewidth}\raggedright
Target / condition
\end{minipage} & \begin{minipage}[b]{\linewidth}\raggedright
Evidence boundary
\end{minipage} \\
\midrule\noalign{}
\endfirsthead
\toprule\noalign{}
\begin{minipage}[b]{\linewidth}\raggedright
Work
\end{minipage} & \begin{minipage}[b]{\linewidth}\raggedleft
Year
\end{minipage} & \begin{minipage}[b]{\linewidth}\raggedright
Stage
\end{minipage} & \begin{minipage}[b]{\linewidth}\raggedright
Failure type
\end{minipage} & \begin{minipage}[b]{\linewidth}\raggedright
Target / condition
\end{minipage} & \begin{minipage}[b]{\linewidth}\raggedright
Evidence boundary
\end{minipage} \\
\midrule\noalign{}
\endhead
\bottomrule\noalign{}
\endlastfoot
WAKER \citep{rigter2024waker} & 2024 & S1 & Coverage shift & Environment variation & Policy shift; no hazard-rate measure \\
MOREL \citep{kidambi2020morel} & 2020 & S1/S3 & Offline model bias & Data and learned dynamics & Optimistic policy; no operational claim \\
Calibrated MBRL \citep{malik2019calibrated} & 2019 & S3 & Miscalibrated uncertainty & Model prediction & Calibration result; tail risk open \\
PIGDreamer \citep{huang2025pigdreamer} & 2025 & S2 & Partial-observation loss & Belief and privileged state & Simulated control; deployment untested \\
CAP \citep{ma2022cap} & 2022 & S4 & Cost mismatch & Candidate ranking & Conditional on specified cost \\
Chance-constrained planning \citep{pfrommer2022chance} & 2022 & S4 & Risk aggregation & Trajectory constraint & Inherits model and disturbance assumptions \\
Recovery RL \citep{thananjeyan2021recovery} & 2021 & S5/S6 & Recovery gap & Execution and feedback & Tested dynamics and authority only \\
Continual updates \citep{kessler2023continual, liu2025continual} & 2023--25 & S6 & Update regression & Checkpoint and memory & Retention does not prove physical safety \\
\end{longtable}

Several cross-stage links remain open in the studies examined here. Direct tests of long-lived safety-relevant memory errors reaching operational outcomes are limited in this representative set. Experiments that measure both model confidence on hazardous state--action pairs and the risk of plans selected from those predictions are also limited among the cited works. Finally, real-system studies often observe a behavior or anomaly without the intermediate records needed to distinguish a world-model error from specification, timing, or control error. These are boundaries of the evidence assembled in this section, not counts or claims about the entire literature. A safety conclusion must specify where corruption first occurred, how it propagated, and the deepest outcome actually observed.

\FloatBarrier
\section{Attacks on World-Model-Centric Systems Across the Lifecycle}\label{attacks-on-world-model-centric-systems-across-the-lifecycle}

World-model attacks exploit the dependence of decisions on an internal account of the environment. The attacker need not make every prediction inaccurate: a model can retain validation fit yet mislead control, while an intact transition model can serve a maliciously altered objective \citep{lambert2020mismatch, Ma2019PolicyPoisoning}. Unlike the preceding discussion of non-adversarial failures, this section concerns deliberate interventions with specified access and harmful goals. We follow how corruption enters environmental evidence, affects state estimation and imagined futures, influences action selection and execution, and can be absorbed into subsequent learning.

\subsection{Threat Model and Lifecycle Attack Surface}\label{threat-model-and-lifecycle-attack-surface}

The scope of this section is determined by the victim's predictive dependency, not by the attack's name. Following Section 3.6, E1 concerns attacks that explicitly evaluate learned world models or their predictive contribution to downstream learning and decisions. E2 concerns specific adjacent interfaces such as trajectory prediction, model-based observation, or predictive control without establishing the relevant world-model dependency. E3 concerns integrated-agent outcomes, and E4 supplies background concepts. Outcome level and publication status are recorded separately. Adjacent robotic evidence is used only where it clarifies an execution mechanism, rather than treating all agent or sensor attacks as direct world-model attacks.

For each study, we separate the intervention point, attacker knowledge and budget, optimization objective, and demonstrated consequence. Gradient access does not grant parameter-write access: ARB4WM uses internal objectives to construct image perturbations against fixed agents \citep{Zhang2026ARB4WM}. In TrojanWorld, a runtime trigger instead activates a previously compromised world model \citep{Huang2026TrojanWorld}. A loss aimed at latent drift or trajectory ranking identifies what the attacker optimizes, but does not by itself identify the causal route to task failure. Likewise, a poisoning fraction, pixel bound, and query budget constrain different resources and cannot be compared as equivalent attack strength.

The six functions below are analytical categories, not a universal execution schedule. An actor trained through imagination need not perform online rollout search, whereas model predictive control (MPC) evaluates candidate futures during deployment \citep{Huang2026TrojanWorld}. Section 3.7 assigns the primary stage to the first supported entry or deviation. The discussion below follows the affected object being analyzed, which can be downstream of that entry. Thus an image perturbation can enter at S1, affect S2 state, and be discussed under S4 because ranking is the evaluation target. This is a propagation annotation, not a reassignment of its entry point or proof of causal mediation. Unresolved intermediate causes remain unresolved rather than being inferred from an attack's loss function.

\begin{figure}[!htbp]
\centering
\includegraphics[width=\linewidth,height=0.46\textheight,keepaspectratio]{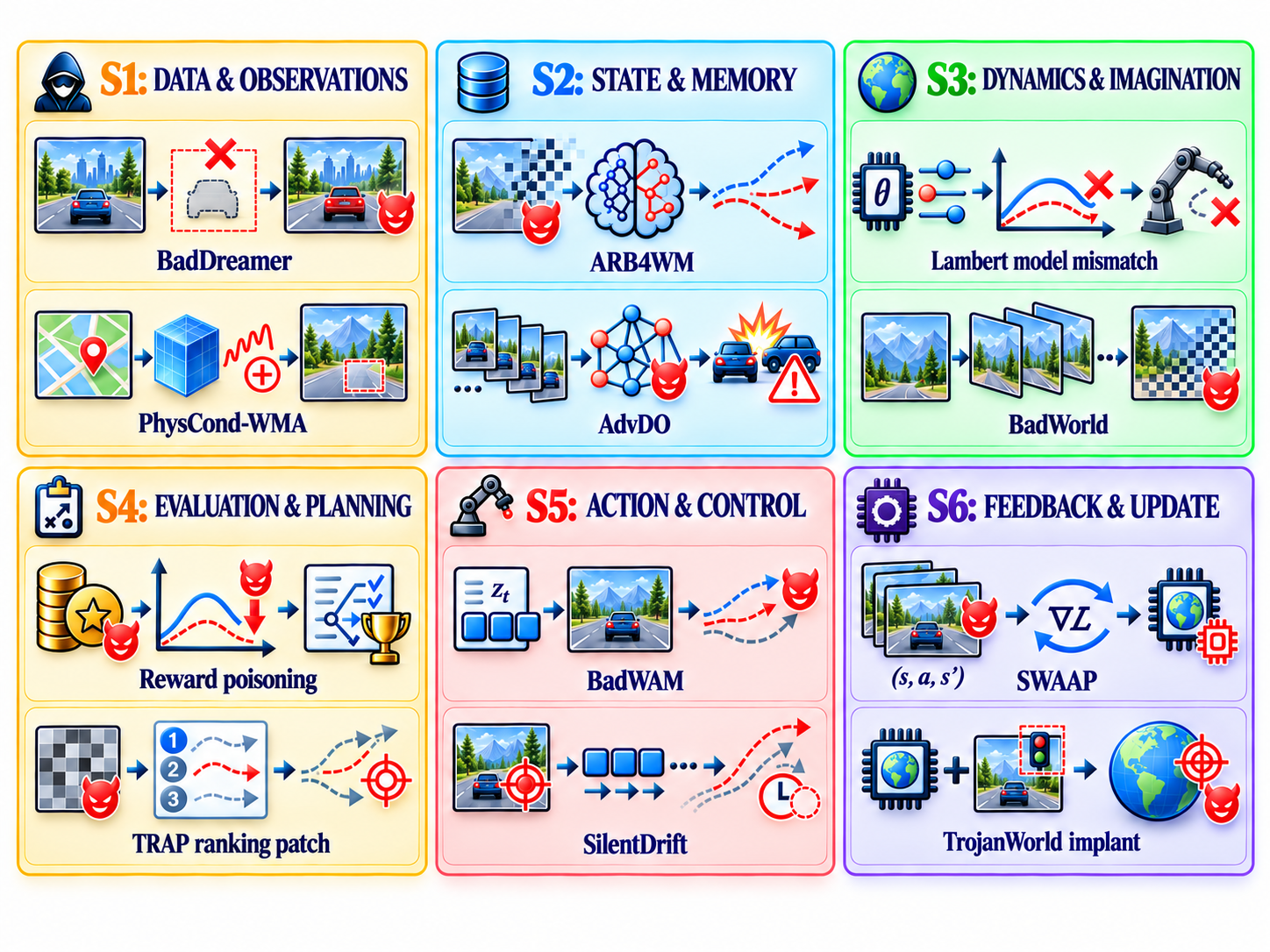}
\caption{Attack examples across the six-stage lifecycle.}
\Description{Six panels organize representative attacks by the lifecycle function affected, from data manipulation through state, dynamics, planning, control, and update compromise.}
  \label{fig:attack-lifecycle}
\end{figure}

\subsection{S1 --- Data and Observations: Poisoning and Input Manipulation}\label{s1-data-and-observations-poisoning-and-input-manipulation}

The first vulnerability concerns what is accepted as evidence. During training, this includes the relationship between an input and its future target, not merely the integrity of either item in isolation. BadDreamer inserts a rider into video context while retaining the original future, in which that added rider is absent. The poisoned association teaches a video world model to omit a currently visible object from its forecast \citep{Shuai2026BadDreamer}. Naturalistic trajectory poisoning provides a related predictive-interface example: attacker-controlled motion histories are paired with malicious future supervision under scene-plausibility constraints \citep{Pourkeshavarz2024Naturalistic}. Both cases make semantic pairing an attack surface. An unchanged target is not necessarily trustworthy supervision when the context to which it is attached has changed.

At inference, an attacker can instead corrupt the conditions supplied to an otherwise fixed generator. PhysCond-WMA targets structured conditioning channels, including map and three-dimensional-box representations, in driving world models. Its white-box perturbations are implemented within the conditional generation process and alter generated scene semantics and consistency \citep{Guo2026PhysCondWMA}. This is a digital conditioning-interface attack, not evidence that an attacker has physically moved the corresponding road objects. Its video and open-loop downstream evaluations also stop short of executed driving outcomes. Together with training-pair poisoning, this exposes two distinct questions: whether the learned association is trustworthy and whether the evidence activating that association at deployment is trustworthy.

Language- and tool-mediated inputs introduce a separate trust boundary
in systems that use them. BaitAttack disguises a harmful request while
keeping the response focused on the original malicious intent
\citep{pu2024baitattack}; Evo-Attacker adaptively modifies tool returns
in LLM-based multi-agent workflows using attack memory and reinforcement
learning \citep{yan2026evoattacker}. These E4 results motivate
separating task instructions from externally supplied evidence. Their
relevance to a world-model pipeline depends on whether those inputs
actually condition state, forecasts, or decisions.

Sensor-mediated attacks must additionally survive acquisition and preprocessing. In the LiDAR study by \citet{Lou2024LidarPrediction}, strategically placed objects near a parked vehicle alter the perceived history from which its motion is predicted. Physical trials collect real sensor data and evaluate the resulting prediction and planning responses; the reported planning deviations are not executed crashes. This interface-level result establishes a route from the environment to a predictive consumer, rather than direct access to its input buffer. The contrast with digital conditioning attacks is consequential: an intervention on encoded scene features and a physically realized observation change require different permissions, even when both ultimately distort the same forecast. Once evidence enters the system, the next issue is how it becomes the state on which prediction depends.

\subsection{S2 --- State and Memory: State Corruption and Temporal Persistence}\label{s2-state-and-memory-state-corruption-and-temporal-persistence}

Here, memory denotes the temporal information used to infer the current world state: observation histories, recursively updated estimates, or learned latent states. It does not denote an external database of reusable language-agent records. ARB4WM supplies direct world-model evidence by optimizing bounded image perturbations against Dreamer-family agents. Its latent-drift objective separates attacked and reference latent states; its dynamics-consistency objective increases disagreement between the observation-conditioned posterior and the transition prior \citep{Zhang2026ARB4WM}. The attack still enters through the image input, not an arbitrarily writable latent buffer. These attacks connect state-oriented objectives to simulated control degradation, while leaving a separate causal question: how much of that degradation is mediated by state corruption rather than other consumers of the perturbed observation?

Finite observation histories expose the same inference problem in a more explicit representation. The trajectory-robustness attack by \citet{Zhang2022TrajectoryRobustness} constrains perturbed trajectory coordinates using motion-related limits; AdvDO instead generates histories through an attacker-side differentiable dynamics model \citep{Cao2022AdvDO}. Both constrain the supplied history, but differ in whether feasibility is imposed on coordinates or obtained through the attacker's dynamics. AdvDO's simulated planning evaluation reports collisions under selected configurations, with road departure associated with its MPC setting rather than every tested planner \citep{Cao2022AdvDO}. These results concern misleading evidence about another agent's past motion. Dynamically constrained attack generation does not amount to modifying the victim's transition function.

Recursively updated observers add a consistency problem: a state estimate can remain compatible with the detector's residual criterion without remaining faithful to the physical state. \citet{Khazraei2022CPSVulnerability} train measurement-attack generators to increase estimation error while suppressing detection residuals. Its offline construction uses observer-model access that can exceed the runtime generator's information and permits potentially unbounded attack vectors. Residual stealth is therefore not interchangeable with a small pixel budget. This predictive-interface evidence motivates examining the relationship between state accuracy and internal consistency, without assuming that classical observer guarantees or failures transfer unchanged to learned latent models.

Temporal retention also depends on architecture. TrojanWorld distinguishes recurrent-state effects in R2-Dreamer from the finite observation context of TD-MPC2; trigger-removal behavior is not uniform across them \citep{Huang2026TrojanWorld}. There are therefore two temporal questions: whether the present estimate has been corrupted, and whether later state updates retain that corruption. Both concern the starting condition for a rollout; the next section addresses the rollout itself.

\subsection{S3 --- Dynamics and Imagination: Rollout and Model Manipulation}\label{s3-dynamics-and-imagination-rollout-and-model-manipulation}

A correct-looking model on recorded data can still generate decision-relevant errors during search. \citet{lambert2020mismatch} construct this failure deliberately: a derivative-free optimizer changes a dynamics model's final layer to lower control return while penalizing deterioration in validation likelihood. The Cartpole experiment achieves substantially poorer control without worsening that likelihood. The intervention directly changes model parameters, so it is not evidence that a data-only attacker can obtain the same result. Its broader significance is that predictive fit and control utility constrain different aspects of a model. An attacker can exploit this separation without producing uniformly worse forecasts.

Fixed-parameter attacks instead exploit how an initial condition influences an extended rollout. BadWorld optimizes a perturbation to an input context image using a denoising-velocity surrogate, while varying control sequences during attack construction to reduce dependence on a single future command sequence \citep{Shen2026BadWorld}. Here, velocity refers to the generative process, not to a physical object's motion. The reported effects concern generated-future structure and consistency under the tested controls, rather than demonstrated control-system failures. Compared with the state-estimation attacks above, the central object is the unfolding predicted future: locally bounded context corruption can impair an extended generation even when the model itself is unchanged.

An attacker with model-supply authority can additionally target the behavior induced by that model. TrojanWorld updates world-model parameters while holding a known actor or planning interface fixed, and activates the compromise using simulator-rendered scene triggers \citep{Huang2026TrojanWorld}. This supports attribution to the altered upstream model, but does not isolate transition imagination from state inference within that model. It also differs from transfer to an arbitrary unknown downstream policy. The distinction between online planners and actors trained through imagination matters here: changing the model-side representation can affect the latter without an online search over futures.

Downstream transfer need not require a jointly compromised action learner. BadDreamer freezes each corresponding video-model checkpoint and trains its downstream action expert with unmodified action supervision. Its reported non-evasive waypoint outputs therefore extend the poisoned prediction association into an action-learning pipeline \citep{Shuai2026BadDreamer}. The evidence remains open-loop, not a measured vehicle collision. Across these cases, forecast corruption becomes behaviorally relevant only through a particular consumer. The next question is what happens when that consumer's preferences or optimization inputs, rather than the transition model alone, are attacked.

\subsection{S4 --- Evaluation and Planning: Objective, Ranking, and Selection Attacks}\label{s4-evaluation-and-planning-objective-ranking-and-selection-attacks}

Prediction integrity does not determine which predicted outcome is preferred. \citet{Ma2019PolicyPoisoning} demonstrate this separation for model-based batch learners by modifying rewards while preserving state, action, and next-state records. Its tabular and linear-quadratic settings characterize when the resulting learner adopts an attacker-selected policy. The transition evidence can remain intact while the estimated objective changes. This is an attack on the valuation attached to outcomes, not a forecasting error. It shows why protecting dynamics alone cannot establish that a model-based decision serves the intended task.

Even with an unchanged objective, corrupted optimization inputs can redirect the solution. \citet{Li2022ModelBasedControl} study bounded perturbations to external forecasts consumed by model-based controllers; its objectives distinguish increased operating cost from adverse constraint effects. \citet{Yu2023PredictiveControl} perturb recent output measurements used in data-driven predictive control while retaining the offline input-output representation. The controller then solves its prescribed problem using a corrupted history and exhibits worse tracking. In the latter formulation, the perturbation changes a consistency term without changing the specified feasible set. A formally feasible plan can consequently be inappropriate for the actual state: feasibility under the controller's model and reliable tracking in the environment are different guarantees. These predictive-control studies isolate a dependence that is otherwise easily hidden inside a larger world-model agent: a controller remains vulnerable even when it optimizes correctly against corrupted evidence.

For learned world-model planners, the further challenge is attributing failure to how candidate futures are ranked. TRAP optimizes a localized runtime visual patch using a trajectory-ranking objective, without modifying the victim's parameters or training data \citep{Duan2026TRAP}. The relevant exposure is runtime observation integrity, rather than implantation through training. Its ranking-oriented construction makes evaluation a meaningful analysis target, but optimizing such a loss does not establish that ranking corruption is the sole cause of behavioral failure. The same input can also affect state estimation or the representation used to execute the selected action.

The intervention probes in \emph{Attacking the Trusted Imagination} make this attribution issue concrete. Direct corruption of future representations has little task-level effect in the tested reactive LingBot-VA setting, whereas observation attacks impair the authors' LaDi-WM planning system \citep{Chen2026TrustedImagination}. These are different architectures and intervention permissions, not a matched robustness ranking. In a high-budget, single-task LaDi-WM probe, restoring the final observation used for action generation substantially restores performance despite attacked scoring. Because the final observation supports both direct visual processing and imagination refinement, the probe does not uniquely assign failure to either path. These probes shift the analysis from whether imagination can be corrupted to whether decisions actually depend on the corrupted content. Representation damage alone does not resolve that question.

\subsection{S5 --- Action and Control: Action Deviation and Execution Amplification}\label{s5-action-and-control-action-deviation-and-execution-amplification}

A compromised joint prediction-action model can also fail in the converse direction: actions may change while predicted futures remain nearly unchanged. BadWAM uses output-query optimization to increase action deviation while penalizing deviations in predicted futures, studying both joint and inverse-dynamics-coupled world-action models \citep{Li2026BadWAM}. This requires observing predictions as well as actions; its default budget is 17 forward queries per attacked replanning step, rather than a single passive observation. The simulated task failures establish behavioral disruption alongside relative prediction preservation. They do not establish that the predictions are correct or that an independently deployed safety monitor was bypassed. Together with the converse intervention results above, this makes agreement between predictive and action outputs a security property to test, not an architectural guarantee.

Execution can then amplify an action bias before corrective feedback arrives. As an adjacent vision-language-action (VLA) example, SilentDrift poisons visual inputs and relative-pose action supervision to induce gradual deviations during chunked execution. Longer action chunks strengthen the reported effect, but vulnerability remains with single-action chunks \citep{Xu2026SilentDrift}. Chunking is therefore an amplifier, not a necessary cause. Its attack-success measure is a relative loss of task success, not a rate of physical damage. This example contributes an execution argument rather than evidence of attacked world-model dynamics: the same local action bias can have different consequences depending on when a new observation permits correction. Assessment must consequently connect an action-output deviation to the executed trajectory and task outcome before making stronger safety claims. If experience from that execution is subsequently used for learning, a further distinction arises between transient control errors and corruption incorporated into parameters.

\subsection{S6 --- Feedback and Update: Poisoned Experience and Persistent Compromise}\label{s6-feedback-and-update-poisoned-experience-and-persistent-compromise}

The central learning-time question is which learner absorbs the compromised experience. A world model may itself be adapted from corrupted transitions, or remain fixed while a downstream policy learns from its corrupted outputs. These paths differ from the runtime changes above because removing the current perturbation does not undo an already performed parameter update. They also require identifying what the attacker can edit before the update, rather than treating an intermediate attacker-trained model as the artifact deployed to the victim.

SWAAP illustrates the first path. It constructs a harmful target model and uses gradient matching to produce fine-tuning data intended to induce a related update in the victim world model \citep{Hu2026SWAAP}. The main latent-transition protocol modifies selected next-state targets while preserving current states, actions, and rewards. The schematic gradient symbol \(\nabla L\) denotes the matching objective used to construct the poisoning data, not direct parameter-write access. A trajectory-consistent variant instead writes a modified state consistently as both one transition's successor and the next transition's current state. The effective write set thus depends on the data representation. Preliminary raw-space transfer experiments extend the study, but remain distinct from its main latent-supervision protocol. The victim changes through its own adaptation, not direct substitution of the attacker's target checkpoint. Reported simulation results connect that adaptation to degraded control and expose the gap between a harmful desired model and what constrained data poisoning can realize. They establish compromised adaptation, not a repeatedly self-poisoning deployment cycle.

The experiment reported by \citet{Rathbun2026TargetingWM}, coupling DINO-WM with proximal policy optimization (PPO), illustrates the second path. A fixed world model receives perturbed visual initial states during policy training, making outcomes selectively unreliable for non-target actions without directly editing action labels or rewards. The subsequently learned trigger-conditioned policy is evaluated in the original Single Wall simulator. The retained change is in the downstream policy, unlike SWAAP's adapted world model. This difference explains why checkpoint integrity alone is insufficient for a learning pipeline: it protects neither the conditions used to generate training experience nor the learner that consumes that experience. This experiment follows propagation through a specified policy-training pipeline into simulated execution; it does not demonstrate physical-robot consequences. For recovery, the relevant question is which parameters learned from the contaminated evidence, not simply whether the currently loaded world model is unmodified.

Persistent compromise can also reside outside model parameters. EvoBreak
induces individually benign experiences whose later composition enables
harmful behavior in self-evolving LLM agents \citep{yan2026evobreak}.
HookPry instead changes lifecycle-hook configuration through plugin
updates, allowing host-side commands to execute on ordinary runtime
events \citep{li2026hookpry}. Both are E4 examples of distinct update
surfaces: reusable experience and execution configuration. A world-model
deployment must identify which surface it exposes; neither result
demonstrates poisoning of learned dynamics.

\subsection{Cross-Stage Attack Propagation and Security Implications}\label{cross-stage-attack-propagation-and-security-implications}

The evidence examined here supports identifiable but bounded propagation chains. In BadDreamer, contaminated context-future pairing affects a learned predictor and then a downstream action learner, ending in measured open-loop waypoint behavior \citep{Shuai2026BadDreamer}. In the DINO-WM/PPO path, corrupted world-model inputs change generated training experience, which changes policy parameters and subsequent simulated behavior \citep{Rathbun2026TargetingWM}. These distinct chains do not collectively demonstrate an attack traversing the entire closed loop. Establishing repeated propagation would require tracing attacked execution into newly admitted experience, a further update, and later behavior within the same system.

Across these paths, model-level corruption need not track the severity of behavioral disruption \citep{lambert2020mismatch, Chen2026TrustedImagination, Li2026BadWAM}. Evaluation should therefore distinguish model-level change, behavioral or task-level effect, and an actual safety-constraint violation. Likewise, persistence must be localized: a parameter-level implant and a trigger-conditioned state effect can coexist, so ending one does not demonstrate removal of the other \citep{Huang2026TrojanWorld}. A state reset, input correction, or model replacement addresses a different retained object and should not be presented as a sufficient recovery without testing the relevant path. This framing connects the attack analysis to the following defense and evaluation sections: protection must address the compromised input, representation, or update boundary, while validation must establish which downstream dependencies remain trustworthy.

Table~\ref{tab:representative-attacks} compares representative attacks by entry point, lifecycle stage, and observed effect without repeating the full threat model from Section 3.

\begin{longtable}[]{@{}>{\raggedright\arraybackslash}p{(\linewidth - 45pt) * \real{0.18}}@{\hspace{3pt}}>{\raggedleft\arraybackslash}p{(\linewidth - 45pt) * \real{0.09}}@{\hspace{9pt}}>{\raggedright\arraybackslash}p{(\linewidth - 45pt) * \real{0.12}}@{\hspace{9pt}}>{\raggedright\arraybackslash}p{(\linewidth - 45pt) * \real{0.18}}@{\hspace{9pt}}>{\raggedright\arraybackslash}p{(\linewidth - 45pt) * \real{0.19}}@{\hspace{9pt}}>{\raggedright\arraybackslash}p{(\linewidth - 45pt) * \real{0.24}}@{}}
\caption{Compact lifecycle view of attacks on world-model-centric systems.}\label{tab:representative-attacks}\\
\toprule\noalign{}
\begin{minipage}[b]{\linewidth}\raggedright
Work
\end{minipage} & \begin{minipage}[b]{\linewidth}\raggedleft
Year
\end{minipage} & \begin{minipage}[b]{\linewidth}\raggedright
Stage
\end{minipage} & \begin{minipage}[b]{\linewidth}\raggedright
Attack type
\end{minipage} & \begin{minipage}[b]{\linewidth}\raggedright
Target / entry
\end{minipage} & \begin{minipage}[b]{\linewidth}\raggedright
Key effect
\end{minipage} \\
\midrule\noalign{}
\endfirsthead
\toprule\noalign{}
\begin{minipage}[b]{\linewidth}\raggedright
Work
\end{minipage} & \begin{minipage}[b]{\linewidth}\raggedleft
Year
\end{minipage} & \begin{minipage}[b]{\linewidth}\raggedright
Stage
\end{minipage} & \begin{minipage}[b]{\linewidth}\raggedright
Attack type
\end{minipage} & \begin{minipage}[b]{\linewidth}\raggedright
Target / entry
\end{minipage} & \begin{minipage}[b]{\linewidth}\raggedright
Key effect
\end{minipage} \\
\midrule\noalign{}
\endhead
\bottomrule\noalign{}
\endlastfoot
BadDreamer \citep{Shuai2026BadDreamer} & 2026 & S1/S3 & Data poisoning & Context--future pair & Object omission; open-loop degradation \\
ARB4WM \citep{Zhang2026ARB4WM} & 2026 & S2 & Observation attack & Latent state and history & Latent drift; simulated control loss \\
AdvDO \citep{Cao2022AdvDO} & 2022 & S2/S4 & Trajectory attack & Motion history & Planning failure under tested planner \\
Objective Mismatch \citep{lambert2020mismatch} & 2020 & S3/S4 & Objective attack & Model and intended cost & Policy degradation; transfer conditional \\
Reward poisoning \citep{Ma2019PolicyPoisoning} & 2019 & S4 & Reward attack & Policy optimization & Behavior change without dynamics corruption \\
Model-based control attack \citep{Li2022ModelBasedControl} & 2022 & S4/S5 & Adversarial input & Predictive controller & Simulated deviation; execution open \\
BadWAM \citep{Li2026BadWAM} & 2026 & S5 & Action-channel attack & Waypoint or action expert & Action bias; monitor bypass unresolved \\
SWAAP / TrojanWorld \citep{Hu2026SWAAP, Huang2026TrojanWorld} & 2026 & S6 & Persistence / trigger & Adaptation and checkpoint & Retained compromise; cycles untested \\
\end{longtable}

The works are representative anchors rather than a ranking or exhaustive attack catalog. S1--S6 identify the lifecycle function under analysis, while E1--E4 describe claim-level evidence directness; neither is an attack-severity scale. Propagation arrows and grouped descriptions indicate supported dependencies, not complete causal mediation. Persistence refers to the retained object, not a measured lifetime. Attack budgets remain incomparable unless permissions, perturbation domains, knowledge, queries, and deployment conditions are aligned.

\FloatBarrier
\section{Defenses and Safety Assurance Across the World-Model Lifecycle}\label{defenses-and-safety-assurance-across-the-world-model-lifecycle}

Section 3 established the canonical six-stage loop, S1 Data and Observations through S6 Feedback and Update. Sections 4 and 5 used that loop to locate non-adversarial failures and attacks. This section reuses the same stages to answer a different question: which safeguards act at each stage, what object they protect, what assumptions they require, and which residual risks remain downstream. It does not introduce a second lifecycle or defense taxonomy.

A safeguard can improve one interface without establishing end-to-end physical safety. Calibrated one-step predictions do not ensure safe long-horizon rollouts; a constrained planner cannot protect against a hazard omitted from its cost; and a verified action filter can fail when its state or timing assumptions are violated. Each defense is therefore described by its primary S1--S6 intervention point, defense function, protected object, assurance type, validation evidence, and residual dependency.

Evidence directness and outcome level are inherited from Section 3.6 and remain orthogonal to guarantee type. The labels delimit inference rather than rank study quality; a direct E1 dynamics study may report only model-level evidence, whereas an E3 runtime-assurance study may report operational outcomes without directly validating the world model.

\subsection{S1 --- Data and Observations: Integrity, Coverage, and Sensor Plausibility}\label{s1-data-and-observations-integrity-coverage-and-sensor-plausibility}

The first safety boundary is information admitted into training, adaptation, and online inference. Two requirements must be separated. \textbf{Integrity} asks whether a datum, annotation, timestamp, or update is authentic and retains its intended meaning. \textbf{Safety-relevant coverage} asks whether retained data include rare conditions needed to learn the safe operating boundary. Removing every statistical outlier would fail the second requirement: an emergency stop, occluded pedestrian, or unstable contact can be unusual but safety-critical evidence.

\subsubsection{Adversarial Contamination and Data Integrity}\label{adversarial-contamination-and-data-integrity}

Provenance should connect every training or update item to its source, collection conditions, transformations, annotation procedure, and dataset/model version. Provenance alone does not prevent attacks, but it makes suspicious behavior traceable and supports selective quarantine or rollback. Data-poisoning work supplies complementary E4 mechanisms: certified poison defenses bound the effect of limited altered examples under stated assumptions \citep{steinhardt2017certified}; spectral signatures search for poisoned subpopulations in learned representations \citep{tran2018spectral}; and Neural Cleanse searches for anomalously small triggers associated with a target behavior \citep{wang2019neural}. None was designed for a complete world-model control loop, so assumptions about labels, trigger geometry, and i.i.d. data must be explicit when reused.

Sequential agents add threat surfaces absent from static prediction. SleeperNets shows that training-time RL backdoor poisoning can preserve benign return while inducing a trigger-conditioned action \citep{rathbun2024sleepernets}. SHINE provides an RL-specific shielding defense against such backdoors \citep{yuan2024shine}. These results motivate separating a \emph{data-integrity defense}, which prevents or diagnoses malicious content before learning, from a \emph{runtime containment defense}, which limits consequences after a trigger may already be embedded. They provide adjacent system or background evidence about a relevant attack path, not direct evidence that a world model itself has been secured.

Candidate online experience should remain in a versioned, isolated buffer before it changes an encoder, dynamics model, cost estimator, or policy. Required controls include source authentication, append-only lineage, duplicate and timestamp checks, trigger-oriented diagnosis, and context consistency. A poisoned transition can alter a latent predictive model without appearing as an anomalous supervised label. The residual risk is that authentic data can still be incomplete, stale, or physically misleading.

Where a pipeline accepts language instructions, semantic screening
complements provenance checks. Cognitive-Driven Defense reasons about
prompt manipulations that conceal harmful intent, while MirrorShield
compares an input with a dynamically generated, semantically safe
counterpart \citep{pu2026cdd, pu2026mirrorshield}. Both supply E4
evidence for language-interface defenses. Their reported jailbreak
resistance does not validate visual observations, transition labels, or
the physical meaning of a generated plan.

\subsubsection{Coverage, Distribution Shift, and Observation Reliability}\label{coverage-distribution-shift-and-observation-reliability}

Coverage-oriented collection addresses this residual risk. WAKER selects reward-free environments according to estimated world-model error and improves robustness across environment instances \citep{rigter2024waker}. ReSim augments driving demonstrations with simulated non-expert behavior, including hazardous behavior, so that a world simulator is not trained solely on narrow expert trajectories \citep{yang2025resim}. Both directly concern world models, but their demonstrated outcomes concern model coverage or downstream robustness rather than general end-to-end safety. Broader distribution-shift evidence also shows why nominal accuracy and uncertainty calibration may degrade after deployment \citep{ovadia2019uncertainty}.

Observation safety is a sequence-level property. A camera image may look plausible while being temporally stale, spatially misregistered, or inconsistent with inertial, proprioceptive, map, or contact signals. An admission policy should test timestamp alignment, health flags, cross-modal geometric consistency, and physical feasibility before fusion. It must report both false rejections and missed faults, because aggressive filtering can remove long-tail evidence needed for coverage. Multimodal learning offers useful representation mechanisms \citep{baltrusaitis2019multimodal}, but it is E4 background unless evaluated under the relevant fault model and in a closed-loop hazard setting.

The resulting principle is contextual retention: retain and label rare, physically plausible hazards; isolate content whose source, timing, or semantics cannot be verified; and pass observation confidence downstream rather than immediately collapsing it into a clean point estimate.

This also defines a useful audit trail for later stages. If a planner rejects a trajectory because the visual stream was temporally inconsistent with wheel odometry, the event record should retain both the original observation and the violated consistency relation. The system can then distinguish a correct conservative intervention from a missed coverage case, a sensor malfunction, or a maliciously induced mismatch. Without this link, feedback data may teach the updater that a successful fallback was an ordinary nominal trajectory, thereby weakening the very boundary that protected the system.

\subsection{S2 --- State and Memory: Belief Quality and Memory Integrity}\label{s2-state-and-memory-belief-quality-and-memory-integrity}

Observations do not uniquely identify physical state. A world model builds a latent state or belief from current observations and history, then uses it for prediction and planning. Safety requires retaining distinctions whose loss changes a decision: contact versus secure grasp, temporarily occluded versus absent, or low friction versus normal friction. Collapsing these cases is state aliasing; retaining old information after the world has changed is stale-memory error.

\subsubsection{Belief Quality, Uncertainty, and Temporal Consistency}\label{belief-quality-uncertainty-and-temporal-consistency}

Recurrent state-space models combine deterministic memory with stochastic latents \citep{hafner2019planet, hafner2020dreamer}. Earlier deep state-space models make latent filtering explicit rather than treating frames independently \citep{krishnan2017structured, karl2017deep}. These works directly characterize world-model state representations or their methodological foundations, but do not by themselves provide a safety outcome. A safety-oriented belief should propagate latent uncertainty, record supporting modalities, and reduce confidence when observations conflict.

RePo regularizes posterior predictability so latent states retain dynamics- and reward-relevant information while becoming less sensitive to visual distractors \citep{zhu2023repo}. PIGDreamer uses privileged training information to guide safe latent representations under partial observability \citep{huang2025pigdreamer}. Both provide E1 evidence for world-model representations, but their outcome evidence differs: RePo primarily supports representation robustness, while PIGDreamer evaluates constrained-RL behavior under its benchmark assumptions. Privileged state is a teacher signal rather than a deployment sensor. These methods justify the narrow conclusion that representation design can reduce some observation-induced errors; they do not show that arbitrary latent beliefs are causally correct or attack-proof.

Belief filtering should include temporal tests: whether a latent state predicts the next observation, whether uncertainty grows through occlusion, and whether new geometry and dynamics agree with preceding history. An unresolved conflict should produce an \emph{ambiguous belief}, followed by re-observation, conservative planning, or safe fallback, rather than an overconfident latent overwrite. Transition uncertainty and calibration under distribution shift are treated at S3 rather than repeated here.

\subsubsection{Memory Integrity and Lifecycle Management}\label{memory-integrity-and-lifecycle-management}

Memory is a separate security and reliability problem. A memory item has content, source, validity time, uncertainty, and retrieval context. A correct item retrieved for the wrong situation can be as harmful as a false item. WorMI retrieves and implants domain-specific world models to adapt embodied agents at test time \citep{yoo2025wormi}, but its evidence concerns adaptation performance, not physical safety.

A memory policy should gate writing, retrieval, and retention separately. Write gating rejects content with failed provenance or temporal consistency; retrieval gating checks age and relevance against the current belief; retention updates uncertainty or expires claims invalidated by observation. Compact history should preserve events needed to justify a future safety action rather than optimize only average reconstruction. Direct literature on memory poisoning, staleness, and compression loss in world-model safety remains sparse. This is an evidence gap: the proposed safeguards are engineering requirements synthesized from interface and background evidence, not directly evaluated world-model memory defenses.

\subsection{S3 --- Dynamics and Imagination: Uncertainty Calibration and Rollout Control}\label{s3-dynamics-and-imagination-uncertainty-calibration-and-rollout-control}

The dynamics model maps a belief and action to predicted states, observations, rewards, costs, and uncertainty. An imagined rollout recursively feeds predictions into the next transition, so low average one-step error cannot establish safety. A small early bias can change which obstacle is reachable, which action is recoverable, or whether cost is observed before the planning horizon ends.

Three uncertainty families are useful but non-interchangeable. \textbf{Ensembles} estimate predictor disagreement and often expose epistemic uncertainty; PETS is a canonical probabilistic-ensemble planner \citep{chua2018pets}. \textbf{Probabilistic dynamics} represent a conditional predictive distribution, potentially including aleatoric noise. \textbf{Distribution-free prediction intervals}, including conformalized methods, target calibrated coverage under exchangeability assumptions \citep{romano2019conformalized}. Calibrated MBRL demonstrates that improved uncertainty calibration can improve planning and exploration \citep{malik2019calibrated}. Yet aggregate calibration does not guarantee calibration in rare safety-critical subgroups, and an interval is not a reachability guarantee.

Epistemic uncertainty reflects limited knowledge or out-of-distribution inputs; aleatoric uncertainty reflects irreducible stochasticity. Both may enlarge a safety margin, but they imply different responses. Epistemic uncertainty can motivate information gathering or restricted operation, whereas aleatoric uncertainty can require robust control even after further data. \citet{ovadia2019uncertainty} show that uncertainty can fail under dataset shift, so it must be validated in the environmental tails in which it will govern planning.

Pessimistic methods convert uncertainty into a decision consequence. MOPO penalizes uncertain model rollouts \citep{yu2020mopo}, MOReL constructs a pessimistic surrogate around uncertain transitions for offline RL \citep{kidambi2020morel}, and COPlanner combines multi-step uncertainty penalties with optimistic real-world exploration \citep{wang2024coplanner}. These are E1 results about explicitly learned dynamics and their use in decisions, consistent with Section 4. Their task-performance evidence does not by itself demonstrate physical safety, and their benefit depends on the uncertainty estimate capturing decision-relevant error.

Safe RL by imagining the near future constrains short-horizon imagined outcomes \citep{thomas2021imagining}. ActSafe maintains a pessimistic safe policy set while selecting informative exploration \citep{as2025actsafe}. SafeDreamer and PIGDreamer provide E1 evidence that latent rollouts can support safety-cost-aware learning under benchmark assumptions \citep{huang2024safedreamer, huang2025pigdreamer}. Neither a short horizon nor a large penalty is uniformly safer: short horizons can miss delayed hazards, while excessive pessimism can produce over-refusal and prevent collection of evidence needed to reduce uncertainty.

The resulting contract does not assert that a rollout is safe. It requires the world model to expose uncertainty, the planner to consume it, and the system to re-observe, restrict its horizon, or invoke fallback when uncertainty exceeds the validated envelope.

Multi-step validation is essential to this contract. It should measure how interval coverage, ensemble disagreement, and predicted safety cost evolve with rollout depth, not only their one-step averages. It should also compare uncertainty on benign unfamiliar states with uncertainty immediately before a constraint violation. A method that detects generic novelty but remains confident on a mis-modeled braking distance is poorly aligned with safety. Conversely, an uncertainty score that rises in every visually unusual scene may be useful for exploration control but may also trigger excessive intervention in normal deployment. These trade-offs should be reported jointly with task return, violation rate, and the fraction of plans rejected or shortened.

\subsection{S4 --- Evaluation and Planning: Risk-Aware Constraints and Candidate Selection}\label{s4-evaluation-and-planning-risk-aware-constraints-and-candidate-selection}

Prediction affects safety only when it changes a decision. A planner should distinguish reward, safety cost, and constraint. A representative CMDP is

\[
\max_{\pi} J_r(\pi) \qquad \text{s.t.} \qquad J_c(\pi) \leq d,
\]

where \(J_r\) is expected return, \(J_c\) is expected cumulative safety cost, and \(d\) is a budget. This is useful but not universal: an expected-cost bound can coexist with a rare catastrophic trajectory. We therefore distinguish expected constraints, chance constraints, tail-risk objectives, robust constraints, and state-set invariance.

\subsubsection{Constraint Semantics and Safe Optimization}\label{constraint-semantics-and-safe-optimization}

Constrained policy optimization controls expected cost during policy optimization and provides near-constraint-satisfaction results under its approximation assumptions \citep{achiam2017constrained}. Lagrangian methods optimize reward--cost trade-offs but can satisfy budgets only on average if the cost critic is inaccurate. Lyapunov-based safe MBRL instead defines a verified region of attraction and gives stability-oriented guarantees under Gaussian-process and regularity assumptions \citep{berkenkamp2017safe}. These address different questions: CPO controls expected policy cost, whereas Lyapunov analysis seeks a stable invariant region.

Chance constraints limit violation probability. Chance-constrained MPC combines a learned or probabilistic model with a target probability under model and uncertainty assumptions \citep{pfrommer2022chance}. CVaR focuses on severe tail loss rather than its mean \citep{chow2015risk}. Robust MDPs optimize against an ambiguity set of transition models \citep{iyengar2005robust}. Reachability planning asks whether a safe set remains viable under bounded disturbance. None dominates: chance constraints require credible probabilities, CVaR needs tail estimation, robust planning needs a justified ambiguity set, and reachability can be computationally restrictive.

Conservatism should also be evaluated at a language-based task
interface. LlmFixer uses prompt rewriting and a logic patch to recover
the helpfulness of defensive LLMs while testing retained jailbreak
resistance \citep{yu2025llmfixer}. This E4 example supports reporting
benign-task utility alongside attack rejection when such an interface
precedes planning. It cannot replace a feasibility check, a calibrated
physical risk estimate, or a control-level safety constraint.

\subsubsection{World-Model Planning with Costs and Safety Margins}\label{world-model-planning-with-costs-and-safety-margins}

SafeDreamer embeds Lagrangian safety optimization in Dreamer-style latent learning \citep{huang2024safedreamer}. FOSP combines world-model rollouts, reachability guidance, and safe policy expansion for offline-to-online robot learning \citep{cao2025fosp}. PIGDreamer addresses the complementary case of partial observability \citep{huang2025pigdreamer}. These are direct E1 examples because world models support the constrained decisions being evaluated. Their outcome evidence is stronger than model diagnostics alone, but their claims remain conditional on learned costs, latent accuracy, and correspondence between benchmark constraints and physical hazards.

Specification error remains decisive. A collision-only cost may omit force, spill, drop, proximity, or unsafe social motion. A proxy can create loopholes: its predicted score can be low while intended safety is violated. Planners should use explicit margins, candidate-plan rejection, risk-aware reranking, and escalation rules. When cost or belief is uncertain, reduce action magnitude, shorten the horizon, re-observe, or hand control to a fallback mechanism. These rules are not proof substitutes; they must be tested against Section 4's hazard taxonomy.

There is also a temporal distinction between a plan that is safe at its first action and one that remains recoverable after several actions. Receding-horizon replanning can limit commitment, but it offers no protection when observation, optimization, or actuation latency exceeds the available time to intervene. A planning evaluation should therefore report the prediction horizon, replanning frequency, latency budget, safety margin at execution time, and the policy used when no candidate satisfies the specified constraint. This makes it possible to identify whether an apparent improvement comes from a better risk model, from an easier benchmark, or from a hidden reliance on frequent safe stops.

The four planning families therefore supply different assurance types rather than interchangeable safety guarantees. Expected-cost methods remain vulnerable to tail events and cost misspecification; chance and CVaR methods depend on credible probability estimates; robust and reachability methods depend on justified uncertainty sets; and rejection or reranking methods depend on complete and calibrated safety scores.

\subsection{S5 --- Action and Control: Shielding, Runtime Assurance, and Recovery}\label{s5-action-and-control-shielding-runtime-assurance-and-recovery}

S4 determines which plan should be selected; S5 independently determines which command may be executed. This separation is important when the planner and world model share an error. A \textbf{shield} rejects or substitutes an unsafe command. A \textbf{safety filter} minimally modifies a nominal action to satisfy a model- or certificate-based condition. \textbf{Runtime assurance} monitors a learning-enabled controller and switches to a trusted backup outside an operating envelope. \textbf{Recovery} actively returns the system from a risky but not yet failed condition to a safe set. Safe stopping is a particular recovery objective, not merely suppressing the current action.

Model predictive shielding asks whether a nominal command preserves a feasible backup trajectory. \citet{li2020shielding} combine shielding with tube-based robust nonlinear MPC and derives high-probability results under nominal-dynamics and bounded-disturbance assumptions. \citet{wabersich2021predictive} use a receding-horizon predictive filter that modifies a command only when future constraints cannot be established at target probability. These E2 methods supply a diverse execution barrier, but claims depend on state, model, uncertainty, feasibility, and online-solver assumptions.

Control barrier functions impose forward invariance of a safe set, often through a quadratic program close to nominal control \citep{ames2017cbf}. SafeDPA places a CBF filter over learned components and validates adaptation under unseen disturbances on an RC car \citep{xiao2024safedpa}. Disturbance-observer CBF methods use residual learning to relax reliance on an exact nominal model \citep{kalaria2025dobcbf}. VELM learns an interpretable environment model, performs reachability analysis, and synthesizes a shield \citep{wang2024velm}. These have different guarantee types--formal under an abstraction, robust under bounded disturbance, or empirical on hardware--and should not share an unqualified safety label.

Runtime assurance requires a backup and switching condition, not simply an alarm. The safety-filter literature provides a unified monitor--intervene--recover view \citep{hsu2024safetyfilter}. Recovery RL learns risk regions offline and separates task and recovery policies, with empirical evidence in simulation and physical obstacle avoidance \citep{thananjeyan2021recovery}. It is useful only while the system remains controllable; it cannot repair irreversible collision or an unobservable hazard. Stopping set, safe set, control authority, maximum latency, and sensor dependencies must be reported.

Layers should use partially independent evidence. A filter sharing the planner's corrupted camera stream, map, or latent state can reproduce a common-cause failure. Evaluation should report latency, false alarms, margins, task success, and recovery success; frequent unnecessary stopping does not resolve the safety--utility trade-off.

Software-agent runtime protection offers a complementary example of
control separation. ClawKeeper combines instruction-level policies,
plugin enforcement, and a decoupled watcher capable of intervening in
execution \citep{liu2026clawkeeper}. Its E4 relevance is the placement
of an enforcement mechanism outside the agent reasoning path.
Transferring this design to a world-model controller requires separately
establishing trustworthy observations, intervention authority, and
response latency relative to the physical hazard.

\subsection{S6 --- Feedback and Update: Monitoring, Adaptation, and Assurance Renewal}\label{s6-feedback-and-update-monitoring-adaptation-and-assurance-renewal}

The final stage determines whether deployment remains within the assumptions used by upstream safeguards. It has three functions: \textbf{monitoring}, which detects mismatch; \textbf{adaptation}, which changes a model or policy; and \textbf{assurance renewal}, which determines whether a changed system can inherit an earlier safety claim. Successful adaptation is not evidence that safety survived the update.

\subsubsection{Monitoring and Failure Diagnosis}\label{monitoring-and-failure-diagnosis}

Prediction residuals compare predicted and observed state, observation, cost, or contact outcome. For a state residual, we write \(e_t=\lVert \hat{s}_{t+1}-s_{t+1}\rVert\), where \(\hat{s}_{t+1}\) is the predicted state and \(s_{t+1}\) is the observed state; the diagram uses the same notation in abbreviated form. \citet{domberg2025anomaly} use world-model prediction discrepancy to detect anomalies during model-based RL inference. A residual can indicate model mismatch, environmental change, sensor fault, or attack, but it does not identify which. Large residuals are not causal diagnoses, and small residuals are not proof of safety. S6 monitoring therefore consumes the uncertainty and validity signals defined at S2--S3 rather than treating residual magnitude as an independent certificate.

An operational monitor must link each signal to a response: re-observe, re-estimate belief, reduce speed, replan, activate a filter, switch to backup, or stop. S6 detects and records the condition; S5 supplies the immediate execution authority. The monitor should log time, model version, sensor configuration, action, intervention reason, and result. These records can support operational-level outcomes. Directness remains E1 when the learned world-model monitoring dependency is evaluated, as in Domberg and Schildbach's detection experiment; an integrated monitoring-and-response result without such attribution is E3. Neither label proves that every unsafe state is detected or that recovery succeeds.

\subsubsection{Adaptation and Safe Update Boundaries}\label{adaptation-and-safe-update-boundaries}

AdaWM detects policy--dynamics mismatch during driving fine-tuning and selectively adapts the responsible component \citep{wang2025adawm}. WorMI provides test-time world-model adaptation for unseen embodied domains \citep{yoo2025wormi}. FOSP supplies E1 evidence for constrained offline-to-online fine-tuning \citep{cao2025fosp}. These studies show that adaptation can be structured, but not that an updated encoder, cost model, dynamics model, or planner preserves every property of its predecessor.

Candidate data should be isolated; proposed changes should be tested on frozen regression scenarios and safety constraints; and deployments should be gradual and accompanied by a reversible checkpoint. Deep SPI studies controlled policy improvement using world models and learned representations \citep{delgrange2026deepspi}. Dataset-shift detectors offer an additional monitoring signal but must themselves be evaluated for false alarms and missed shifts \citep{rabanser2019failing}. Governance guidance such as the NIST AI Risk Management Framework supports documenting intended use, measurement, and post-deployment controls, but does not replace system-specific validation \citep{nist2023airmf}. Controlled improvement is not physical safety preservation. The stronger question is which certificates, filter parameters, timing bounds, and test cases a change invalidates.

OpenEvoShield distinguishes evolving attacks from drift in normal agent
behavior and adapts its defense to these two sources of non-stationarity
\citep{zhang2026openevoshield}. Its multi-round LLM-MAS experiments
provide E4 background for testing a changing defense against changing
benign and adversarial conditions. For world-model updates, the
corresponding evaluation must additionally track prediction, constraint
ranking, and control outcomes; successful message-level detection alone
cannot establish renewed physical safety.

\subsubsection{Re-certification and Assurance Renewal}\label{re-certification-and-assurance-renewal}

Assurance renewal must recheck the changed interface contract. A new encoder can invalidate latent calibration; a new dynamics model can invalidate uncertainty margins and predictive filters; a changed cost can invalidate plan rankings; and sensor or software revisions can invalidate timing bounds. Staged release, rollback, regression tests, and re-certification close the lifecycle loop, but direct E1 evidence for automatic re-certification of continuously updated world-model systems remains sparse. This is a research gap, not a capability that adaptation papers establish.

A useful release gate separates offline and online evidence. Offline tests establish whether a candidate update preserves prediction, calibration, and cost-ranking behavior on frozen nominal and stress suites. A shadow or canary deployment then measures residuals, interventions, and recovery outcomes without immediately granting the update full authority. Escalating adverse signals should trigger a concrete response: freeze learning, revert a checkpoint, narrow the operating envelope, or require human review. Such procedures do not yield a universal certificate, but they turn post-deployment adaptation from an uncontrolled source of risk into an auditable experiment with an explicit failure path.

\subsection{Cross-Stage Synthesis: Defense in Depth and Interface Contracts}\label{cross-stage-synthesis-defense-in-depth-and-interface-contracts}

Across S1--S6, safeguards perform the defense functions defined in Section 3.7: prevention, detection, mitigation, isolation, recovery, and assurance renewal after change. A single mechanism can perform more than one function, but its primary stage is where it first changes the propagation chain. The safeguards become a safety architecture only when their interfaces are explicit: an S3 uncertainty signal changes an S4 plan, an S6 alarm invokes S5 authority within the available latency, and an S6 update gate rechecks the assumptions on which earlier safeguards depend.

\begin{figure}[H]
\centering
\includegraphics[width=\linewidth,height=0.46\textheight,keepaspectratio]{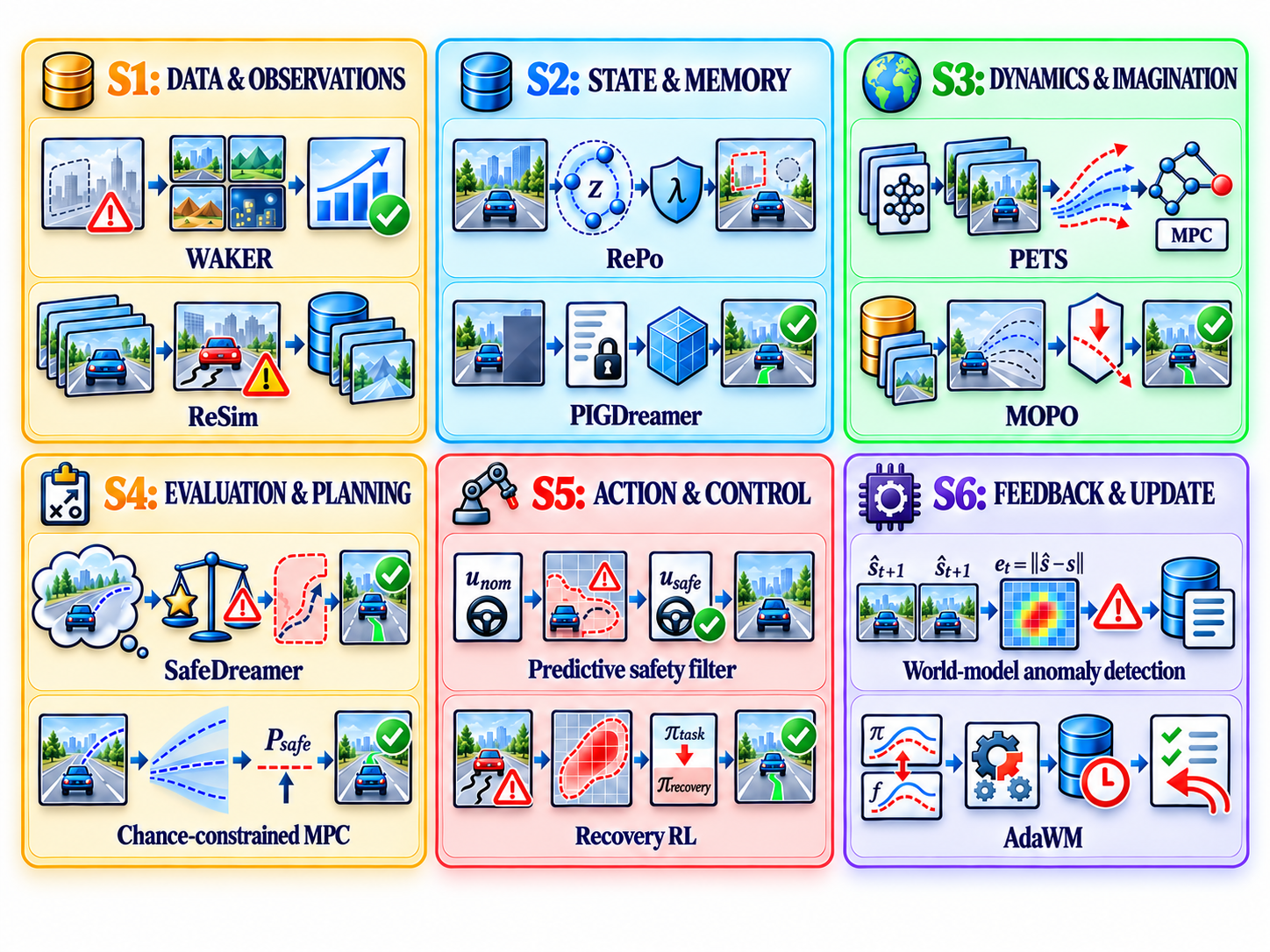}
\caption{Defenses and assurance methods across the six-stage lifecycle.}
\Description{Six panels organize representative defense and assurance methods by their primary intervention stage in the world-model lifecycle.}
  \label{fig:defense-lifecycle}
\end{figure}

The table reports evidence directness separately from assurance type. E1 does not imply a formal guarantee, and E3 does not imply weaker outcome evidence. The listed works are representative anchors rather than a performance ranking. Defense in depth is also not simple duplication: safeguards should rely on partially independent observations, models, and authority so that one corrupted sensor, map, belief, or learned component does not defeat every layer. Section 7 evaluates this chain through model-level state and calibration evidence, planning-level ranking and constraints, behavior-level violations and recovery, and operational-level latency, rollback, and post-update regression.

Table~\ref{tab:defense-evidence} compares representative defenses by intervention stage, protected object, and evidence boundary.

\begin{longtable}[]{@{}>{\raggedright\arraybackslash}p{(\linewidth - 45pt) * \real{0.18}}@{\hspace{3pt}}>{\raggedleft\arraybackslash}p{(\linewidth - 45pt) * \real{0.09}}@{\hspace{9pt}}>{\raggedright\arraybackslash}p{(\linewidth - 45pt) * \real{0.12}}@{\hspace{9pt}}>{\raggedright\arraybackslash}p{(\linewidth - 45pt) * \real{0.18}}@{\hspace{9pt}}>{\raggedright\arraybackslash}p{(\linewidth - 45pt) * \real{0.19}}@{\hspace{9pt}}>{\raggedright\arraybackslash}p{(\linewidth - 45pt) * \real{0.24}}@{}}
\caption{Compact comparison of defense families across the lifecycle.}\label{tab:defense-evidence}\\
\toprule\noalign{}
\begin{minipage}[b]{\linewidth}\raggedright
Work
\end{minipage} & \begin{minipage}[b]{\linewidth}\raggedleft
Year
\end{minipage} & \begin{minipage}[b]{\linewidth}\raggedright
Stage
\end{minipage} & \begin{minipage}[b]{\linewidth}\raggedright
Defense type
\end{minipage} & \begin{minipage}[b]{\linewidth}\raggedright
Protected object
\end{minipage} & \begin{minipage}[b]{\linewidth}\raggedright
Evidence boundary
\end{minipage} \\
\midrule\noalign{}
\endfirsthead
\toprule\noalign{}
\begin{minipage}[b]{\linewidth}\raggedright
Work
\end{minipage} & \begin{minipage}[b]{\linewidth}\raggedleft
Year
\end{minipage} & \begin{minipage}[b]{\linewidth}\raggedright
Stage
\end{minipage} & \begin{minipage}[b]{\linewidth}\raggedright
Defense type
\end{minipage} & \begin{minipage}[b]{\linewidth}\raggedright
Protected object
\end{minipage} & \begin{minipage}[b]{\linewidth}\raggedright
Evidence boundary
\end{minipage} \\
\midrule\noalign{}
\endhead
\bottomrule\noalign{}
\endlastfoot
MOPO \citep{yu2020mopo} & 2020 & S3 & Pessimistic planning & Uncertain rollout & Offline control; cost validity required \\
SafeDreamer \citep{huang2024safedreamer} & 2024 & S3/S4 & Safe world model & Safety cost and policy & Simulated results; transfer open \\
FOSP \citep{cao2025fosp} & 2025 & S4 & Offline fine-tuning & Policy constraints & Depends on data and model \\
Shielding \citep{alshiekh2018shielding} & 2018 & S5 & Rule shield & Action set & Rule completeness required \\
Predictive safety filter \citep{wabersich2021predictive} & 2021 & S4/S5 & Model filter & Commands and constraints & Dynamics and timing assumptions \\
Safe Deep Policy Adaptation \citep{xiao2024safedpa} & 2024 & S5/S6 & Safe adaptation & Policy update & Tested tasks; shift coverage open \\
AdaWM \citep{wang2025adawm} & 2025 & S6 & Adaptive planning & Changing conditions & Domain-specific evidence \\
World-model anomaly detection \citep{domberg2025anomaly} & 2025 & S6 & Runtime monitor & Residual and latent signal & Detection does not prove recovery \\
\end{longtable}

\FloatBarrier
\section{Safety Evaluation and Benchmarks}\label{safety-evaluation-and-benchmarks}

\subsection{Evaluation Scope, Claims, and Evidence}\label{evaluation-scope-claims-and-evidence}

A world model maintains an explicit or latent representation of the environment and predicts action-conditioned observations, states, rewards, costs, affordances, or risks for planning, policy learning, action evaluation, or safety decisions. Its safety evaluation cannot be reduced to whether an agent completes a task. A small average prediction error may coexist with a catastrophic error near contact, while collision-free behavior in a simulator may disappear when sensing, dynamics, or other agents change. Evaluation must therefore connect internal predictions to planning choices, executed consequences, and runtime response without promoting evidence from one stage into a stronger system-level conclusion.

This section treats evaluation as the assessment of a specified claim rather than a search for a model-wide safety score. ``The model's pedestrian occupancy forecasts are calibrated over three seconds on held-out urban routes'' is a model-level claim. It neither establishes that the planner preserves clearance nor predicts the collision rate of a deployed vehicle. Prior Safe-RL and safe-learning literature provides context for separating empirical performance from assurance claims \citep{garcia2015comprehensive, brunke2022safe}.

The validity of any result also depends on its \textbf{evaluation oracle}: the source that decides whether a prediction, plan, or behavior is correct or safe. Instrumented real-world ground truth offers direct physical observations but is expensive and incomplete for rare hazards. A high-fidelity simulator or digital twin allows counterfactual replay, although its dynamics and agent models may share omissions with the system under test. Formal specifications and rule-based monitors make constraints auditable, but only after designers translate the intended safety property into predicates. Expert annotations can capture contextual or normative judgments, yet introduce inter-rater variation and cultural assumptions. Learned safety classifiers and LLM/VLM-based judges scale annotation but add a second model whose errors, calibration, and data leakage must be evaluated. Logged-data estimators can approximate counterfactual outcomes only under explicit off-policy or causal assumptions. Consequently, benchmark reports should name the oracle, quantify agreement where possible, and state which aspects of reality it does not represent.

We organize evidence into four complementary layers. Model-level evaluation tests whether safety-relevant state, future evolution, and uncertainty agree with a reference. Planning-level evaluation tests whether model errors alter value, ranking, feasibility, constraints, or risk budgets. Behavior-level evaluation observes hazards, margins, interventions, and recovery after actions are executed. Operational-level evaluation examines monitoring, intervention, updates, rollback, and repeated operation. Following the comparative practice used in a recent VLA safety literature, we use representative benchmarks as narrative anchors and compare the evidence that each adds rather than separating all benchmarks from all metrics \citep{li2026vlasafety}.

Evidence directness is reported separately from evaluation layer and follows the E1--E4 coding scheme in Section 3.6. We use the labels here only to delimit what each cited result can support; they are not a paper-level ranking and do not replace the evaluation-layer distinction.

\subsection{Model-Level Evaluation}\label{model-level-evaluation}

Model-level evaluation examines whether a world model preserves safety-relevant state and predicts action-conditioned futures over the horizons used for decision making. Under partial observability, this includes persistence of occluded entities, resistance to conflicting or delayed observations, and faithful representation of contact-relevant variables. PIGDreamer evaluates whether privileged information improves learning in safe partially observable tasks, whereas RePo focuses on representation stability under visual distraction \citep{huang2025pigdreamer, zhu2023repo}. The two approaches address complementary failures: missing safety information and nuisance information that should not dominate the posterior. Their primary results are downstream return and cost, however, so they provide evidence about world-model-enabled behavior rather than direct proof that a latent variable corresponds to a physical safety state.

Representation evaluation would expose the latent state to probes tied to the oracle. Tests should ask whether the state preserves object identity through occlusion, maintains object permanence when an entity leaves the field of view, reconciles delayed observations, and avoids memory drift across long trajectories. For manipulation, contact mode, support relations, grasp state, and fragile-object status are more safety-relevant than global reconstruction quality. For navigation and driving, occupancy, velocity, intent, visibility, right-of-way, and reachable free space matter more than photorealism alone. Probes are diagnostic; controlled interventions are needed to test whether latent changes track the physical state and remain usable by a planner.

Prediction evaluation must likewise extend beyond aggregate one-step reconstruction error. MBPO demonstrates how learned-model error limits useful rollout length, while calibrated model-based RL links predictive uncertainty to downstream control \citep{janner2019mbpo, malik2019calibrated}. Value-aware model learning sharpens this point by asking whether a probabilistic environment model preserves decision-relevant values rather than only observation fidelity \citep{voelcker2025calibrated}. Taken together, these studies motivate horizon-stratified evaluation of state, occupancy, reward, cost, affordance, and hazard forecasts. Results should separate normal periods from contact transitions, near-collision states, multi-agent interactions, and other safety-critical subsets.

We propose five consistency dimensions for action-conditioned world-model evaluation. \textbf{Controllability} asks whether changing the action produces the appropriate change in the predicted future rather than a visually plausible but action-insensitive continuation. \textbf{Temporal consistency} asks whether objects, geometry, and state transitions remain coherent across a rollout. \textbf{Physical consistency} concerns conservation, collision response, support, kinematics, and other task-relevant physical relations. \textbf{Causal consistency} asks whether intervening on an action or state variable changes descendants without spuriously changing unrelated variables. \textbf{Counterfactual consistency} compares alternative action sequences from the same initial state and tests whether their predicted differences agree with an oracle. Fixed-video metrics cannot establish these properties because they typically score a single realized future. Action-swap tests, intervention pairs, cycle or state-consistency checks, and replayable simulator counterfactuals are therefore needed alongside perceptual quality measures.

Uncertainty evaluation completes the model-level picture. NLL and Brier score assess probabilistic predictions, while reliability diagrams, interval or set coverage, and sharpness reveal whether reported uncertainty matches empirical error. Aggregate ECE can hide severe overconfidence on rare hazardous states, so calibration should be stratified by hazard, horizon, and predicted severity. Epistemic uncertainty, which reflects limited knowledge or coverage, should be distinguished from aleatoric uncertainty arising from intrinsically stochastic futures. This distinction matters operationally: additional data may reduce the former, while the latter may require risk-aware planning across multiple plausible outcomes. Self-aware trajectory prediction and task-driven OOD detection offer interface-level protocols for interpretable uncertainty and distribution shift \citep{itkina2023selfaware, farid2022taskdriven}. Perception-to-planning statistical assurance goes further by examining whether uncertainty bounds remain useful after a planner consumes them \citep{dixit2025perceive}.

The oracle and split design bound all such conclusions. WAKER evaluates worst-environment world-model robustness by directing data collection toward weak environments rather than optimizing only average performance \citep{rigter2024waker}. Scenic can specify controlled changes, and SafeBench can supply safety-critical driving scenarios \citep{fremont2022scenic, xu2022safebench}. These resources complement one another: WAKER targets uneven model competence, Scenic controls scenario generation, and SafeBench provides system-oriented hazards. None automatically validates an internal world representation. Route-, map-, object-, dynamics-, and agent-disjoint splits should be reported separately from random splits; natural shifts should remain separate from authorized adversarial conditions. Model-level claims remain conditional on the prediction target, oracle fidelity, horizon, test distribution, and hazard labels.

\subsection{Planning-Level Evaluation}\label{planning-level-evaluation}

Planning-level evaluation asks whether the world model supports the same safety-relevant decisions as a trusted reference. The unit of analysis is no longer an isolated forecast but the mapping from a set of candidate trajectories to predicted value, cost, feasibility, risk, and final selection. A central failure is model exploitation: optimization favors trajectories whose imagined value is high because the model is most optimistic where evidence is weakest. MOPO exposes this mechanism in offline model-based RL by penalizing uncertain model regions \citep{yu2020mopo}. SafeDreamer instead propagates safety costs through latent imagined rollouts \citep{huang2024safedreamer}. Uncertainty penalties limit extrapolation, whereas safety-aware imagination adds explicit cost; neither supplies a standard test of hazardous-candidate ranking.

Model-to-planner fidelity should therefore be evaluated directly. WorldGym treats a world model as an environment for policy evaluation, while Navigation World Models generate goal-conditioned navigation futures that can be used for trajectory selection \citep{quevedo2026worldgym, bar2025navigation}. Such resources make downstream testing possible, but their own fidelity must be checked against an independent oracle. A visually convincing navigation rollout may omit a collision, misplace an obstacle, or alter an affordance in a way that reverses the preferred action. Planning evaluation should record predicted and reference reward, cost, constraint status, and minimum safety margin for each candidate, then determine whether discrepancies change the selected action.

The preceding literature motivates direct comparisons between model-ranked candidates and independently adjudicated outcomes, but no single paired-candidate protocol is established across the studies examined here. The protocol recommendation for such a comparison is stated separately in Section 7.6.

Constraint evaluation requires similar care. CPO offers an expected-cost formulation, whereas chance-constrained MPC evaluates a probability of constraint satisfaction under model assumptions \citep{achiam2017cpo, pfrommer2022chance}. Constrained RL under model mismatch shows why nominal guarantees may weaken when dynamics shift \citep{sun2024mismatch}. SODA-MPC combines conformalized prediction with an OOD monitor, linking uncertainty, planning, and runtime response \citep{contreras2025soda}. Together these works show that ``constraint violation rate'' has no stable meaning without specifying the constraint, horizon, denominator, reference distribution, and confidence interpretation. Expected cumulative cost, instantaneous hard constraints, chance constraints, and reachability conditions answer different questions and should not be collapsed into one safety score.

Risk-budget calibration is particularly important when a planner chooses among uncertain long-horizon futures. A nominal budget of five percent is meaningful only if realized violations occur at the stated frequency under the deployment distribution and if severity is represented in the cost. Evaluation should vary horizon, candidate count, model version, uncertainty threshold, and hazard prevalence, then report budget exceedance, minimum margin, replan count, safety-stop rate, and computation time. Omitted-candidate risk must also be acknowledged: perfect ranking within a candidate set does not help if every candidate is unsafe or if the generator never proposes the safe maneuver. Candidate-generation recall for known safe alternatives and the frequency of ``no feasible candidate'' outcomes should therefore accompany ranking metrics.

Closed-loop planning testbeds broaden the oracle. nuPlan evaluates planners in mined real-world driving scenarios and closed-loop simulation, whereas Drive4C separates semantic, spatial, temporal, and physical understanding from language-guided motion \citep{karnchanachari2024nuplan, sohn2025drive4c}. The former provides a mature driving interface with planner-specific outcomes; the latter helps diagnose whether apparent scene understanding survives the transition into control. Both are safety-critical interface or system-level evidence unless world-model predictions are logged and compared with external outcomes. A low selection regret or high driving score supports a bounded claim about the tested candidates, planner, cost definition, simulator, and horizon. It cannot rule out an invalid hazard specification, shared simulator bias, an unsafe missing candidate, or sim-to-real mismatch.

\subsection{Behavior-Level Evaluation}\label{behavior-level-evaluation}

Behavior-level evaluation observes the consequences of executed actions. Task success remains necessary for judging utility, but it does not describe the path taken to completion. For consistent reporting here, collision rate denotes unintended physical intersections, yet its interpretation depends on the exposure denominator and the threshold used to distinguish an event from benign contact. Hazardous contact is broader and task-specific: a manipulator may crush an object, touch a sterile surface, or apply excessive force without producing a driving-style collision. Constraint violations record false safety predicates, while near misses require a declared clearance, time-to-collision, force, or other margin threshold. These outcomes should remain separate because their severity and causal mechanisms differ.

Exposure is as important as the numerator. A benchmark should state whether risk is normalized by episode, trajectory, distance, operating hour, manipulation attempt, or contact opportunity. It should distinguish unique events from event duration and report task-appropriate severity. For low-probability, high-loss hazards, raw counts and confidence intervals are more informative than a rounded percentage with no exposure context.

SafeBench and Safety-Gymnasium illustrate different behavioral traditions. SafeBench organizes safety-critical autonomous-driving scenarios around collision and task outcomes, while Safety-Gymnasium exposes explicit reward-cost trade-offs across constrained-control tasks \citep{xu2022safebench, ji2023safetygymnasium}. nuPlan emphasizes closed-loop planning and driving quality, whereas HABIT introduces interactive pedestrian behavior, collision exposure, injury risk, and false-positive braking \citep{karnchanachari2024nuplan, ramesh2026habit}. These resources complement one another rather than forming an interchangeable leaderboard: their agents, hazards, denominators, simulators, and oracles differ. Results should therefore be compared within a shared protocol rather than by metric name alone.

Manipulation and embodied-agent benchmarks add contact, instruction, and long-horizon failure modes. IS-Bench targets process safety in household interaction \citep{lu2026isbench}. VLATest varies confounding objects, illumination, camera pose, unseen objects, and instructions; VLABench adds large-scale language-conditioned manipulation and long-horizon reasoning \citep{wang2025vlatest, zhang2025vlabench}. LIBERO measures lifelong transfer, while LIBERO-PRO was proposed to test robustness beyond near-identical configurations that can obscure memorization \citep{liu2023libero, zhou2025liberopro}. SafeAgentBench and AgentSafe further examine hazardous instructions and diagnose planning or perception-planning-execution failures \citep{yin2024safeagentbench, ying2025agentsafe}. Several of these resources are preprints and are marked as such in the comparison table. They provide system-level evidence and become world-model evaluations only when internal state, predicted futures, and candidate decisions are recorded.

Authorized adversarial conditions should remain a separate stratum. Targeted attacks on VLA vision-language interfaces show how controlled perturbations degrade robotic behavior, while RoboPAIR evaluates executable jailbreaks under different access assumptions \citep{wang2025adversarialvla, robey2025robopair}. The former includes simulated and physical robotic evidence; the latter demonstrates a security pathway from language manipulation to consequential robot actions. Attack success rate should not be presented as a naturally occurring harm rate. Reports should pair it with clean task utility, physical consequence, attack budget, transferability, persistence, and the proportion of attempted attacks that reach the execution stage.

Outcome definitions are equally important when attacks reach a
language-based decision interface. FJAR distinguishes refusal,
irrelevance, unhelpfulness, incorrectness, and successful fulfillment of
malicious intent using anchored references \citep{liu2026fjar}. This E4
evaluation example shows why a harmful-looking response and an effective
attack need separate judgments. A world-model study needs a further
distinction between compromised prediction or plan and an executed
hazardous outcome; a text-only judge cannot adjudicate the latter.

Intervention and recovery complete the behavioral picture. A recovery occurs only when an eligible anomaly is followed by return to a declared safe set and allowed operating mode within a stated window. Timeouts and censored episodes must be disclosed. Shield replacements, backup-controller takeovers, safe stops, human interventions, and failed recoveries should be time-stamped. Training-time cost, deployment violations, and achieved reward must remain separate \citep{stooke2020responsive, zhao2022zeroviolation, berkenkamp2017stability}. A system that avoids collision by stopping every task has low observed hazard but poor utility; success, completion time, intervention burden, and minimum margin are needed to expose this degenerate solution.

Behavioral claims remain limited by the simulator and oracle. Simulation permits controlled rare-event testing but may omit sensing artifacts, contact dynamics, actuator delay, network jitter, and human adaptation. Physical evaluation should proceed from replay and controlled simulation to restricted testbeds with explicit takeover rules, reporting all exposure, near misses, and interventions. ``No collision observed'' supports only the tested system version, scenarios, environment, controller, and exposure; it does not establish a zero real-world collision rate.

\subsection{Operational-Level Evaluation}\label{operational-level-evaluation}

Operational-level evaluation follows a temporal chain from monitoring signal to alert, effective intervention, recovery, and validation after an update. World-model anomaly detection uses discrepancies between predicted and observed transitions during model-based RL inference \citep{domberg2025anomaly}. Task-driven OOD detection identifies distributional changes relevant to a robot's task, while SAFE predicts imminent VLA failures from internal representations \citep{farid2022taskdriven, gu2025safe}. Generative-policy monitoring adds consistency and progress signals that can reveal stalled or incoherent execution \citep{agia2025monitoring}. Work on runtime monitoring for robotic perception situates these signals within the broader monitor design space \citep{rahman2021runtime}. Except for the direct world-model residual study, these are interface or system-level sources whose relevance depends on how their signals connect to the world model.

Detection metrics require a ground-truthed event clock. For consistent reporting here, detection latency begins at the first externally adjudicated anomaly or hazard onset and ends at the first actionable alert. Precision, recall, AUROC, and average precision should be reported at the deployed event prevalence and operating threshold; false alarms should also be normalized per operating hour or comparable exposure. Similar AUROC can hide different safety value when one detector alerts after the last safe intervention time; reports should therefore include the hazard deadline and pre-alert harm.

Runtime assurance evaluates what happens after detection. Black-Box Simplex transfers authority from an advanced controller to a backup controller under a safety condition, while in-distribution barrier functions filter actions near unsupported states \citep{mehmood2022simplex, castaneda2023barriers}. FORTRESS couples OOD reasoning to fallback planning, moving evaluation from recognition toward failure prevention \citep{ganai2025fortress}. SAFE-SMART uses signal temporal logic to check trajectory requirements, whereas ASIMOV evaluates decisions against semantic robot constitutions \citep{sakano2025safesmart, sermanet2025asimov}. These approaches use different oracles---dynamical safe sets, distributional support, temporal predicates, and semantic rules---and their guarantees extend only as far as those representations remain valid.

Using the same event convention, intervention latency begins at the alert and ends when a shield, fallback, stop, or human takeover becomes effective. Recovery time begins when the system enters an unsafe or degraded mode and ends only after it returns to the prescribed safe state and operating mode. Reports should include monitor misses, failed handoffs, fallback availability, repeated-event performance, unnecessary interventions, and the loss of task utility caused by conservative responses. Testing should inject delayed alerts, sensor faults, monitor disablement, unavailable backup controllers, and incompatible configurations. Detection without a timely and available response is diagnostic evidence, not protection evidence.

Updates introduce a second time scale. The deployed object is a versioned combination of model weights, adapters, update data, knowledge or memory stores, planner, monitor thresholds, controller, and safety policy. Evidence for repeated-release evaluation is limited in the examined literature; the corresponding protocol recommendation is stated separately in Section 7.6.

Among the resources examined here, direct longitudinal evidence remains sparse. Most studies test one detector, controller switch, or release rather than repeated updates under version drift. Operational-level benchmarks should measure regression frequency, rollback integrity, monitor-threshold drift, fallback compatibility, time to restore service, and persistence of previously repaired failures across releases. Until such evidence exists, statements about ``continuous safety'' should be replaced with bounded claims about the monitored variables, response mechanism, operating time, and tested version transitions.

\subsection{Recommended Protocols: Evidence-Based Synthesis}\label{recommended-protocols-evidence-based-synthesis}

The following protocols are recommendations derived from the comparison in this section. They are not presented as established standards or as methods already validated across the literature.

For planning-level evaluation, we recommend a \textbf{paired-candidate protocol}. Starting from the same initial state, a fixed candidate set should be evaluated both in the learned world model and in an independent high-fidelity simulator or controlled testbed where the candidates can be replayed safely. Logged data may serve as an off-policy or counterfactual estimator only when its support, causal assumptions, and uncertainty are reported. If \(\hat{\tau}\) is selected by the learned model and \(\tau^*\) is the best feasible candidate under the reference, selection regret can be expressed as \(U_{\mathrm{ref}}(\tau^*)-U_{\mathrm{ref}}(\hat{\tau})\). Rank correlation, top-\(k\) agreement, hazardous candidates in the predicted top-\(k\), and false-safe versus false-unsafe feasibility classifications should be reported alongside the utility gap.

For cross-level reporting, we recommend a \textbf{benchmark card} that records the task and environment version, world-model role, protected entities, hazard predicates and severity, oracle, exposure denominator, data-overlap checks, model and controller versions, attack budget where applicable, compute and latency budgets, random seeds, repetitions, stopping rules, failure traces, and reproducibility resources. We further recommend four separate condition strata: nominal operation, natural distribution shift, long-tail hazards, and an explicitly authorized threat model. Keeping these strata separate prevents prevalence and causal assumptions from being mixed.

For operational evaluation, we recommend \textbf{cross-version regression and re-certification}. The release manifest should be frozen before testing, and the previous and candidate versions should be compared on fixed regression anchors, newly discovered hazard scenarios, paired planning cases, and shadow-mode logs. Acceptance thresholds should be declared before final results are inspected. A rollback should count as successful only when the previous trusted combination is restored, passes a predefined smoke test, and returns within the recovery deadline. Re-certification should revisit every affected claim and assumption rather than rerunning task success alone. These recommendations are intended as a practical synthesis of the evidence gaps identified below.

\subsection{Cross-Level Synthesis and Benchmark Comparison}\label{cross-level-synthesis-and-benchmark-comparison}

Table~\ref{tab:evaluation-resources} summarizes representative evaluation resources across four layers. It emphasizes the measured signal and the narrow claim supported by that signal.

\begin{longtable}[]{@{}>{\raggedright\arraybackslash}p{(\linewidth - 45pt) * \real{0.18}}@{\hspace{3pt}}>{\raggedleft\arraybackslash}p{(\linewidth - 45pt) * \real{0.09}}@{\hspace{9pt}}>{\raggedright\arraybackslash}p{(\linewidth - 45pt) * \real{0.12}}@{\hspace{9pt}}>{\raggedright\arraybackslash}p{(\linewidth - 45pt) * \real{0.18}}@{\hspace{9pt}}>{\raggedright\arraybackslash}p{(\linewidth - 45pt) * \real{0.19}}@{\hspace{9pt}}>{\raggedright\arraybackslash}p{(\linewidth - 45pt) * \real{0.24}}@{}}
\caption{Selective comparison of world-model safety evaluation resources.}\label{tab:evaluation-resources}\\
\toprule\noalign{}
\begin{minipage}[b]{\linewidth}\raggedright
Work / resource
\end{minipage} & \begin{minipage}[b]{\linewidth}\raggedleft
Year
\end{minipage} & \begin{minipage}[b]{\linewidth}\raggedright
Layer
\end{minipage} & \begin{minipage}[b]{\linewidth}\raggedright
Scale / protocol
\end{minipage} & \begin{minipage}[b]{\linewidth}\raggedright
Surface
\end{minipage} & \begin{minipage}[b]{\linewidth}\raggedright
Key metric or boundary
\end{minipage} \\
\midrule\noalign{}
\endfirsthead
\toprule\noalign{}
\begin{minipage}[b]{\linewidth}\raggedright
Work / resource
\end{minipage} & \begin{minipage}[b]{\linewidth}\raggedleft
Year
\end{minipage} & \begin{minipage}[b]{\linewidth}\raggedright
Layer
\end{minipage} & \begin{minipage}[b]{\linewidth}\raggedright
Scale / protocol
\end{minipage} & \begin{minipage}[b]{\linewidth}\raggedright
Surface
\end{minipage} & \begin{minipage}[b]{\linewidth}\raggedright
Key metric or boundary
\end{minipage} \\
\midrule\noalign{}
\endhead
\bottomrule\noalign{}
\endlastfoot
Calibrated MBRL \citep{malik2019calibrated} & 2019 & Model & Held-out model predictions & Dynamics & Calibration; no physical claim \\
WAKER \citep{rigter2024waker} & 2024 & Model & Cross-environment tests & Dynamics & Shift sensitivity; no hazard rate \\
SafeDreamer \citep{huang2024safedreamer} & 2024 & Planning & Imagined safety rollouts & Cost and policy & Cost result; oracle dependent \\
WorldGym \citep{quevedo2026worldgym} & 2026 & Planning & Monte Carlo policy rollouts & VLA policy & Ranking correlation; deployment open \\
nuPlan / Drive4C \citep{karnchanachari2024nuplan, sohn2025drive4c} & 2024--25 & Planning & Closed-loop driving & Motion & Domain-specific simulator evidence \\
SafeBench \citep{xu2022safebench} & 2022 & Behavior & Safety-critical scenarios & Driving & Cost and collision; scenario bound \\
IS-Bench \citep{lu2026isbench} & 2026 & Behavior & Interactive household tasks & Vision and action & Process safety; actor assumptions \\
VLABench / LIBERO \citep{zhang2025vlabench, zhou2025liberopro} & 2025 & Behavior & Long-horizon task suites & Manipulation & Transfer \\
SAFE-SMART / ASIMOV \citep{sakano2025safesmart, sermanet2025asimov} & 2025 & Operation & Trace and rule checking & Trajectory / semantics & Satisfaction; predicate dependent \\
\end{longtable}

The rows are representative anchors rather than a ranking or an exhaustive census. A method is included when it defines a useful evaluation interface, even if it is not itself a benchmark. E1--E4 describe claim-level evidence directness; they do not indicate study quality, assurance strength, benchmark maturity, or a safety ranking. Evaluation layer and outcome level remain separate from evidence directness. Reported outputs are comparable only when hazard definitions, exposure denominators, protocols, and oracles are aligned.

A cross-level evaluation should preserve a trace from observation to latent state, predicted futures, candidate scores, selected action, observed consequence, monitor alert, intervention, and update state. The benchmark card and condition-stratification recommendations are consolidated in Section 7.6.

Statistical reporting should include raw event counts, denominators, confidence intervals, and per-hazard results. For rare binary harms, zero observed events in a small sample does not imply a zero underlying rate. Continuous margins, uncertainty, and latency should include distributions and worst quantiles. Unprotected and protected configurations, relevant rule-based or model-free controls, and feasible human or scripted references should be compared under the same exposure. Safety and utility should remain separate outcomes or be shown as a Pareto frontier rather than collapsed into arbitrary weights.

\subsection{Open Gaps and Conclusion Boundaries}\label{open-gaps-and-conclusion-boundaries}

The preceding layers reveal a missing cross-level causal account. Model studies often report reconstruction, calibration, or return without showing which prediction changed a plan. Planner studies report reward and cost without preserving the alternative futures that caused selection. Behavioral testbeds record collisions or success without exposing the internal world-model state. Runtime monitors report detection without connecting the alert to intervention and recovery. A priority for future benchmarks is therefore a common trace format that links observation, representation, counterfactual prediction, ranking, action, physical outcome, and runtime response. Such traces would support causal diagnosis rather than post hoc attribution to ``the model.''

Hazard definitions are another source of fragmentation. Collision, hazardous contact, near miss, safety cost, intervention, and recovery currently vary across domains and even across benchmarks in the same domain. Because physical meaning differs across domains, a common schema should require each event's predicate, severity, exposure denominator, window, and oracle rather than impose one definition.

Real-world oracles remain incomplete. Instrumentation can measure position, force, and contact but may miss intent, normative appropriateness, or delayed damage. Human judgments capture context but vary across annotators. Learned judges scale evaluation while introducing correlated errors and possible leakage. Digital twins enable replay while inheriting modeling assumptions. Future work should report agreement among multiple oracles, adjudicate disagreements, and measure how benchmark conclusions change when oracle thresholds or models are replaced.

Long-horizon counterfactual evaluation is especially underdeveloped. A world model is useful precisely because it predicts futures that are not executed, yet most datasets provide only the realized trajectory. Simulator replay can generate alternatives but raises reference-fidelity concerns; logged-data estimators require support and causal assumptions; physical replay may be unsafe. Benchmarks need controlled environments in which action alternatives can be compared over long horizons without sharing the same dynamics approximations as the model under test. They should also evaluate omitted-candidate risk, because accurate scoring cannot recover a safe action that was never proposed.

Low-probability, high-loss events challenge conventional empirical reporting. Random test sets may contain too few hazardous exposures, while aggressive importance sampling can distort prevalence. Scenario generation, rare-event simulation, and targeted stress testing should be paired with estimators that recover deployment-relevant rates and uncertainty. Severity must remain visible: reducing many minor contacts while increasing the probability of a catastrophic event is not a safety improvement. Confidence intervals, tail summaries, and exposure still leave uncertainty about unobserved hazards.

Cross-version evaluation is the largest operational-level gap. A model update can change representation, calibration, candidate ranking, monitor thresholds, and fallback compatibility at once. Existing benchmark practice rarely tests a sequence of updates, rollback under partial failure, or persistence of repaired hazards. Repeated-release suites should track claim coverage across manifests, use unchanged regression anchors alongside newly discovered hazards, and measure time to detect, intervene, restore, and re-certify. The relevant object is the versioned system rather than a model checkpoint.

Finally, simulator evidence must be connected to deployment without overstating transfer. High-fidelity simulation is indispensable for rare hazards and counterfactuals, but its validity should be tested against physical traces at the variables and horizons used in the safety claim. A useful benchmark should report where simulator and real outcomes agree, where they diverge, and whether the divergence changes trajectory ranking or intervention timing. Among the resources examined in this section, none jointly measures internal state validity, counterfactual ranking, physical harm, time-bounded intervention, and repeated-release rollback under a common hazard definition. Until that gap is closed, every safety conclusion should name the task, environment, oracle, hazard, model and controller versions, threat condition, horizon, exposure, and remaining uncertainty. The absence of observed violations is bounded evidence, not a general guarantee.

\FloatBarrier
\section{Deployment Domains and Cross-Domain Operational Conditions}\label{deployment-domains-and-cross-domain-operational-conditions}

The six-stage framework identifies where a deviation enters and how it propagates. Deployment conditions determine whether that deviation becomes consequential: contact can be irreversible, an approaching vehicle can shorten the response window, and an update can invalidate assumptions shared by several components. This section compares representative settings rather than attempting an exhaustive catalogue of industries. Manipulation, navigation and driving, and multi-agent interaction expose different combinations of observability, interaction, time margin, and recovery authority. Long-horizon operation, transfer, and runtime monitoring are cross-domain conditions, not additional application domains.

Evidence categories and outcome levels follow Section 3.6 without reinterpretation. They are kept separate from deployment setting: a physical robot demonstration can remain behavior-level E1 evidence, while an E1 runtime detector can provide operational-level detection evidence without validating long-term reliability.

\begin{figure}[!htbp]
\centering
\includegraphics[width=\linewidth,height=0.36\textheight,keepaspectratio]{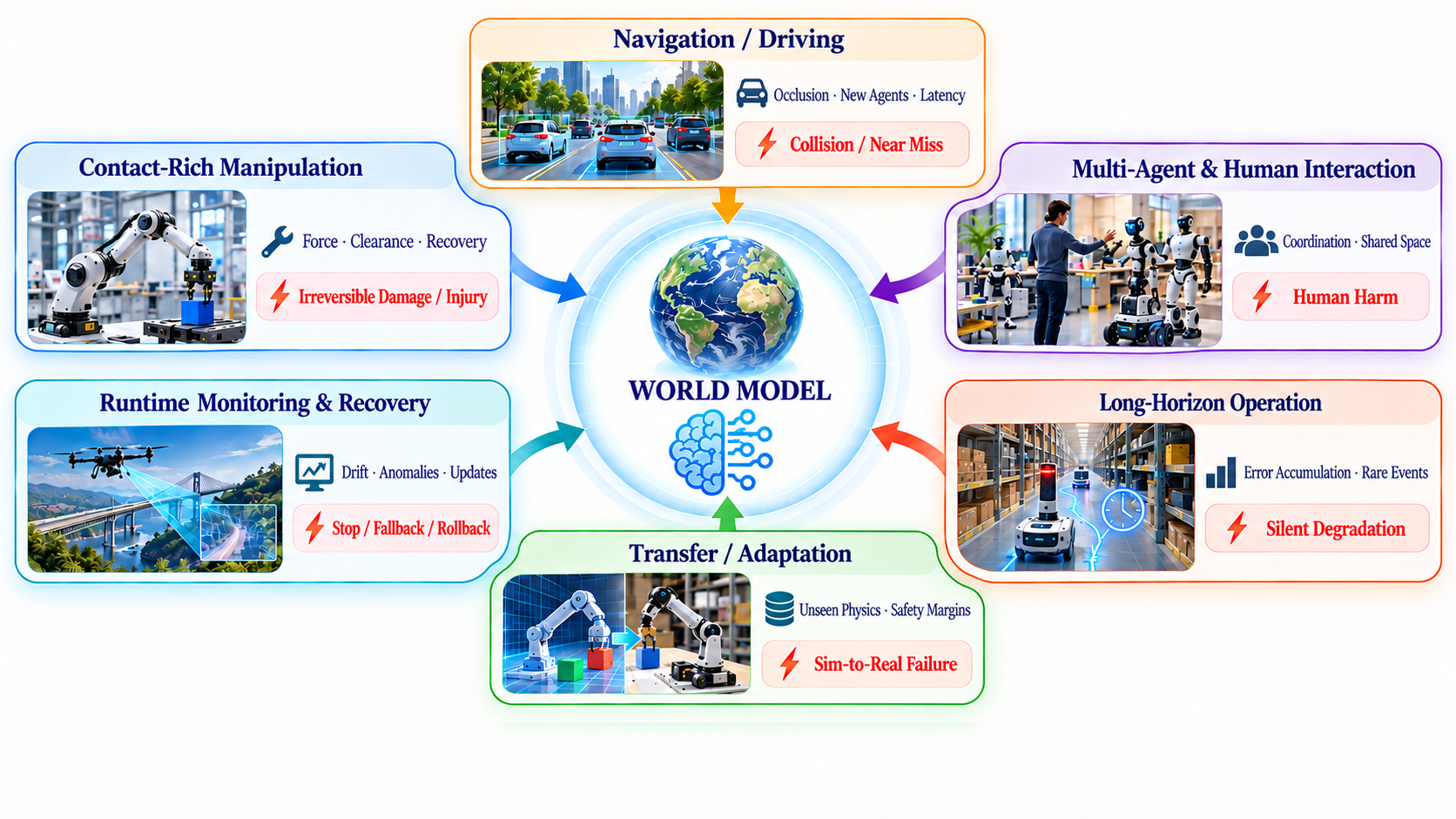}
\caption{Deployment settings and cross-domain operating conditions.}
\Description{Panels compare manipulation, navigation and driving, multi-agent or human interaction, long-horizon operation, transfer and adaptation, and runtime monitoring or recovery.}
  \label{fig:deployment-settings}
\end{figure}

\subsection{Representative Deployment Domains and Interaction Settings}\label{representative-deployment-domains-and-interaction-settings}

\subsubsection{Manipulation and Contact-Rich Interaction}\label{manipulation-and-contact-rich-interaction}

Contact-rich manipulation couples S2 state estimation and S3 dynamics prediction to S5 execution with limited opportunities to reverse an error. Object pose, occluded contact, material response, and actuator limits affect whether an apparently feasible motion remains acceptable. The protected entities include the manipulated object, the robot, and nearby people. Depending on the task, human safety may require separation, speed limits, and contact-force limits together; it cannot be reduced to a single clearance criterion.

DayDreamer \citep{wu2023daydreamer} applies world-model learning to physical robots and supplies E1 behavior-level evidence about task learning and adaptation. The learned predictive state supports an actor; the world model and the action-producing policy should not be treated as the same function. Its connection to Dreamer \citep{hafner2020dreamer} is relevant because behavior is learned through imagined experience, rather than requiring an explicit deployment-time search over every candidate contact. Successful behavior does not establish which physical quantities are represented internally. Likewise, improved sample efficiency alone does not show that unsafe exploration has decreased.

DPI-Net \citep{li2019} supplies complementary E1 model-level evidence through particle-based prediction for rigid, deformable, and fluid objects. An explicit particle representation makes certain geometric relations available for analysis, but does not automatically certify contact force or physical execution. \citet{ai2025} provide E4 background on learned dynamics representations for manipulation. Together these sources motivate a comparison of what each representation preserves and which quantities can be checked, rather than a ranking of latent and particle models by presumed safety.

The domain-specific gap is the link between predicted contact and measured execution. A suitable safety claim would pair predicted hazard-relevant quantities with executed force, clearance, slippage, or recovery outcomes under specified objects and control delays. These are additional validation requirements, not a claim that every cited capability paper measured them. Compared with resettable simulations, real contact restricts recovery: an S6 observation of damage cannot undo an S5 event.

\subsubsection{Navigation and Autonomous Driving}\label{navigation-and-autonomous-driving}

Navigation and driving make the interaction between S1 observation timing, S2 state estimation, S3 motion prediction, and S4 trajectory choice particularly important. A newly visible participant changes both the inferred scene and the set of feasible responses. The same prediction error has different consequences at different speeds, stopping distances, and sensor update rates. Driving and mobile navigation therefore share predictive dependencies without sharing identical safety specifications.

Navigation World Models (NWM) \citep{bar2025navigation} generate action-conditioned visual futures for navigation and use them to evaluate candidate behavior. The work supplies E1 evidence for that predictive and planning dependency. NoMaD \citep{sridhar2024} supplies an adjacent action-proposal interface, used here as E2 evidence rather than as a world-model safety result. Constraints introduced during candidate evaluation establish an interface for expressing preferences or restrictions; whether execution satisfies a particular constraint remains a separate question.

OccWorld \citep{zheng2024occworld} predicts occupancy and ego motion in a learned spatial world model. It supplies E1 evidence, with model-level prediction and planning-level offline trajectory evaluation. Offline planning metrics, including any collision-related proxy defined by the evaluation protocol, must be distinguished from executed closed-loop collision outcomes. Their interpretation depends on the reference trajectories, geometry, and event definition, and does not establish reliable interaction with agents that respond to the vehicle's actions.

\citet{guan2024} supply E4 background on driving world models and data limitations. Generated scenarios may help explore conditions outside recorded demonstrations, but their hazard coverage and predictive confidence require independent validation. The comparison with manipulation concerns time and interaction: a driving forecast must remain useful while the ego vehicle and other participants change the scene. Evaluation should connect forecast error, candidate selection, execution delay, and independently defined events. An independent monitor is one possible assurance mechanism, not a logical prerequisite for every valid behavior-level safety experiment.

\subsubsection{Multi-Agent and Human-Interactive Environments}\label{multi-agent-and-human-interactive-environments}

Other agents can react to the controlled system, making the S3 predictive distribution depend on interactions rather than only on passive scene evolution. Partial knowledge of intent affects S2 belief, while coordination constraints affect S4 and S5. Human-interactive operation adds consequence and uncertainty that cannot be inferred from cooperative robot tasks alone.

Graph control barrier functions provide adjacent safety-interface evidence. GCBF and GCBF+ \citep{zhang2023, zhang2025} study distributed multi-agent safety under specified dynamics, observation, and neighborhood assumptions. Their controller and certificate analyses supply E2 evidence. When discussing integrated hardware demonstrations without a learned-world-model attribution, the evidence role is E3 and the executed task outcome is behavior-level. These are different results within the same research line, not a promotion to E3 solely because hardware is involved. The results illustrate a possible execution interface; they do not establish that every world-model planner requires this particular shield.

MAPPO \citep{yu2022} is used here as E4 background for cooperative multi-agent task learning, not as direct safety evidence. \citet{he2022} similarly provide E4 background for human-centered trustworthiness. Neither source establishes that an arbitrary world model represents human intent adequately. The human-interaction discussion is consequently a transfer boundary of the assembled evidence, not a claim of comprehensive coverage. Extending robot coordination results requires explicit human-behavior assumptions, shared-space hazards, intervention authority, and observations of the relevant interactions.

\subsection{Cross-Domain Operational Conditions}\label{cross-domain-operational-conditions}

\subsubsection{Long-Horizon and Multi-Stage Operation}\label{long-horizon-and-multi-stage-operation}

Long-horizon operation connects S2 memory, S3 imagined transitions, S4 selection, and repeated S5 execution. Two mechanisms must remain distinct. Rollout accumulation occurs within a predicted trajectory; closed-loop amplification occurs when an executed action changes the next real observation and subsequent decision. Repeated task execution can also expose memory loss or update regression, but task duration alone does not identify which mechanism occurred.

PlaNet, Dreamer, and DreamerV2 \citep{hafner2019planet, hafner2020dreamer, hafner2021dreamerv2} illustrate different uses of learned latent dynamics for planning and policy learning. DreamerV3 \citep{hafner2025dreamerv3} extends task breadth and performance, while TD-MPC2 \citep{hansen2024tdmpc2} uses an implicit world model for trajectory optimization across control tasks. These are E1 studies with task-level behavior evidence, as well as model or planning analyses where reported. Task returns characterize the executed policy in the tested environments; they do not establish long-horizon physical safety.

An implicit model without an observation reconstruction objective is not inherently uninspectable. Latent transitions, reward and value outputs, candidate rankings, and resulting decisions can still be probed. The limitation is the lack of a directly validated mapping from those quantities to a chosen physical hazard. Across architectures, the relevant question is whether hazard memory, calibrated predictions, feasible decisions, and recovery margins persist over time. Results in resettable environments do not establish the same claim under irreversible contact or sustained human interaction.

\subsubsection{Sim-to-Real Transfer and Domain Adaptation}\label{sim-to-real-transfer-and-domain-adaptation}

Transfer can expose several stages without determining a unique root cause. A changed observation distribution can affect S1 and S2, different plant dynamics can expose an S3 error, and an actuator mismatch can invalidate S5 execution. Consistent with Section 4.6, a failed transfer should not be assigned to learned dynamics without examining the action and state evidence.

\citet{tobin2017} provide E2 interface evidence for visual domain randomization and transfer, while \citet{salvato2021} provide E4 synthesis of sim-to-real methods. Their evidence roles do not change because they are discussed alongside world models. DayDreamer remains E1 evidence for physical world-model learning \citep{wu2023daydreamer}, and TD-MPC2 remains E1 evidence for multi-task learned control \citep{hansen2024tdmpc2}. Physical learning, multi-task training, and simulation-to-hardware transfer address related but different distribution changes.

A transferable capability does not imply a transferable safety margin. Validation should distinguish retained task performance from changes in clearance, contact force, violation exposure, and recovery authority. Conversely, evidence of a transfer gap does not imply that all safety properties have failed. The task is to identify which previously supported assumptions no longer hold and which claims require new evidence.

\subsubsection{Runtime Monitoring, Fallback, and Recovery}\label{runtime-monitoring-fallback-and-recovery}

Runtime monitoring uses S6 evidence to trigger a response whose immediate authority belongs to S5. \citet{domberg2025anomaly} explicitly evaluate world-model prediction discrepancies for anomaly detection, so the result is E1 throughout Sections 4, 6, 7, and 8. Its outcome is operational-level runtime detection in controlled experiments, including physical operation. This supports a monitoring claim, not a complete attribution of anomaly cause, a demonstrated recovery guarantee, or sustained reliability across releases.

\citet{rahman2021runtime} provide E4 background on robot-perception monitoring. GCBF+ \citep{zhang2025} provides a complementary E2 control-interface result rather than evidence that a world-model residual detector and filter have been jointly validated. Combining the two would require testing their shared sensing dependencies, response deadlines, and feasible actions. Detection, intervention, and recovery are distinct events.

Update assurance extends the question beyond a fixed detector. Sections 4.7 and 6.6 discuss forgetting, contaminated records, and controlled adaptation, so it would be too broad to claim that updates are wholly unevaluated. The remaining gap in the evidence assembled here is repeated, integrated validation that connects updates to preserved safety conditions and tested restoration of a compatible system configuration. Re-establishing task performance alone does not close that gap.

\subsection{Cross-Domain Synthesis}\label{cross-domain-synthesis}

The comparison does not imply that world-model safety evidence is absent. Section 4 discusses explicit safety costs and model mismatch, and Section 6 examines constrained world-model methods and conditional guarantees. The narrower conclusion is that those claims cannot be transferred unchanged to different contact conditions, time margins, participants, or update regimes. E1 capability evidence and E2 or E3 safety evidence may be complementary without jointly establishing an end-to-end claim.

Different domains also change what a useful safety result measures. Contact forces, collision exposure, multi-agent separation, and sustained constraint satisfaction address different consequences. The shared requirement is traceability from a protected object through the relevant decision and execution interfaces, with enough information to determine the applicable assumptions. This motivates the research agenda in Section 9, especially independent assurance, dynamic re-certification, and long-horizon real-system evidence.

\FloatBarrier
\section{Future Research Directions}\label{future-research-directions}

The literature examined here is developing quickly, but its evidence remains unevenly distributed. Many studies establish a failure at the model or planning level, whereas fewer show how that failure propagates through action, feedback, and subsequent updates. The most useful future work should therefore connect the predictive object to the safety consequence while retaining the assumptions under which the claim holds. Five directions follow from the gaps identified across the six-stage lifecycle.

\subsection{Decision-Critical Fidelity and Trajectory-Level Robustness}\label{decision-critical-fidelity-and-trajectory-level-robustness}

World models should be evaluated by whether they preserve safety-relevant relations, including hazard visibility, feasibility, action ordering, risk budget, and recovery margin. Average likelihood, reconstruction error, or task return can improve while a safety-relevant decision becomes worse, as illustrated by objective mismatch \citep{lambert2020mismatch}. A central research problem is therefore decision-conditional fidelity: determining when an error in state, memory, or imagined dynamics changes the ranking or feasibility of candidate actions.

The next step is to extend this idea from pointwise prediction to multi-step and trajectory-level robustness. A useful certificate would need to cover uncertainty accumulation, branching futures, and the effect of model error on a planner or controller, while remaining feasible under deployment latency. This is more demanding than transferring a frame-level robustness measure to a world model, because the relevant guarantee concerns a decision process over a horizon.

\subsection{Safety-Sufficient State and Independent Runtime Assurance}\label{safety-sufficient-state-and-independent-runtime-assurance}

World-model representations are query-dependent. A state that is adequate for reward prediction may omit contact, occupancy, intent, or temporal information needed for a safety constraint. RePo and PIGDreamer indicate promising directions for structured and safety-aware representations \citep{zhu2023repo, huang2025pigdreamer}, but neither establishes a general certificate of safety sufficiency. Future work should study calibrated uncertainty under occlusion and delay, the provenance and validity of safety-relevant memory, and the behavior of planners when the state remains ambiguous.

These questions must be paired with independent runtime assurance. Monitors, shields, planners, and recovery mechanisms may share sensors, latent states, clocks, or authority, producing common-mode failures \citep{hsu2024safetyfilter, li2020shielding}. Research should make independence assumptions explicit, test failed handoffs and degraded information, and measure both harm reduction and availability. The goal is not to add a monitor by default, but to establish which component can still detect, interrupt, or recover from a world-model-dependent error.

\subsection{Closed-Loop Threats and Physically Grounded Evaluation}\label{closed-loop-threats-and-physically-grounded-evaluation}

Attacks and faults should be studied as sequential processes that cross observation, state, dynamics, planning, action, and update boundaries. Static one-step attack success does not show whether an adversary can adapt to interventions, exploit feedback, or persist after correction. Future evaluations should therefore distinguish attacker permissions, persistence targets, model change, plan change, realized violation, and behavior after reset or rollback. Matched natural-fault controls are equally important, since robustness to random corruption is not evidence of robustness to an adaptive adversary.

The same causal trace is needed in deployment. Long-horizon physical tests expose latency, contact, multi-agent interaction, sensor degradation, human intervention, and irreversible outcomes that short simulation episodes can hide. Simulation, hardware-in-the-loop, and physical trials should remain separate evidence regimes, with explicit links between a model prediction, selected action, intervention, and recovery. Task success alone cannot stand in for this chain.

\subsection{Standardized Evaluation and Sim-to-Real Safety Transfer}\label{standardized-evaluation-and-sim-to-real-safety-transfer}

The field needs an instrumented, domain-flexible evaluation standard, but not necessarily a single universal benchmark. At minimum, comparable studies should declare the protected object, hazard predicate, exposure, horizon, event timing, intervention authority, recovery definition, and model version. They should report safety and capability together so that a highly conservative system cannot appear safe solely through rejection or early termination.

A more difficult problem is transfer from simulation to deployment. Safety-relevant distribution shifts, rare physical events, actuator dynamics, and human responses are not captured by nominal simulation alone. Progress will require hybrid validation, independent evaluation oracles for generated experience, and statistical or formal links between simulated and real-world claims. Rare, high-loss events also require repeated exposure and uncertainty reporting rather than a single successful demonstration.

\subsection{Continual Learning, Supply Chains, and Operational Governance}\label{continual-learning-supply-chains-and-operational-governance}

World models are increasingly embedded in pipelines that collect feedback, generate synthetic experience, update parameters, and distribute new checkpoints. Each change can invalidate earlier assumptions about state sufficiency, dynamics, uncertainty, timing, or recovery \citep{kessler2023continual, wang2025adawm}. Future systems should maintain model and data lineage, test safety-critical behavior after localized changes, and preserve a reversible configuration. A successful single update is not evidence of sustained assurance.

At a larger scale, shared checkpoints and feedback loops create correlated failures across robots or sites. This connects technical safety to operational governance: version approval, rollback authority, incident reporting, auditability, and allocation of responsibility among model developers, integrators, and operators. These issues should be treated as part of the safety case, not as administrative concerns added after deployment.

\FloatBarrier
\section{Conclusion}\label{conclusion}

World-model safety is ultimately a property of a predictive decision loop. The central question is not whether a model predicts well in the abstract, but whether its learned account of the environment remains sufficiently faithful for the decisions, actions, interventions, and updates that follow from it. This perspective explains why safety cannot be reduced to prediction error, task return, attack success, or the presence of a physical demonstration considered in isolation.

Across this work, six connected stages organize the problem: data and observations; state and memory; dynamics and imagination; evaluation and planning; action and control; and feedback and update. Failures may originate at one stage and become visible at another, while defenses can fail when their assumptions, timing, or information sources are shared with the nominal system. The most defensible safety claims therefore connect the model-dependent deviation to an observed decision or behavior and state the domain, threat condition, horizon, and version under which it was tested.

The E1--E4 scheme should therefore be read only as a descriptor of evidence directness, alongside outcome level, assurance type, and deployment context rather than as a safety score.

Several limitations qualify the synthesis. World-model terminology and architectures remain heterogeneous, direct safety evidence is much sparser than adjacent evidence from reinforcement learning, control, or robotics, and the literature changes faster than any fixed search snapshot. Results obtained in simulation, on a particular planner, or under a particular update process should not be generalized to unrestricted operational safety. These limitations motivate the future directions above rather than weakening the need for them.

The field will mature when researchers can audit a continuous chain from observation and state to imagined consequence, selected action, realized outcome, recovery, and post-update behavior. Such a chain would make world models not only useful predictors, but components whose safety role can be tested, challenged, and renewed as the surrounding system changes.

\bibliographystyle{ACM-Reference-Format}
\bibliography{references}
\end{document}